\documentclass{iucr} 

\journalcode{J}

\usepackage{makecell}
\usepackage{amssymb}
\usepackage{amsmath}
\usepackage{mathrsfs}
\usepackage{epsfig}
\usepackage{graphicx}
\usepackage{fancybox}

\usepackage{cases,bm,multirow}
\usepackage{url,xcolor}
\usepackage{mwe,tikz}
\usepackage[percent]{overpic}

\usepackage{enumitem}
\usepackage{comment}

\newtheorem{rem}{ Remark}[section]
\graphicspath{{images/}}
\newenvironment{sumMy}[1][]{\textstyle\sum}

\begin{document}

%\title{Iterative hyperspectral x-ray ptychography with (partially) known dictionary }
%\title{Iterative thickness reconstruction for  x-ray spectroscopic ptychography with (partially) known dictionary }
%\title{Iterative chemical mapping for  x-ray spectroscopic ptychography with (incomplete) dictionary }
\title{Iterative X-ray Spectroscopic Ptychography}

\ifx \aff \undefined
\author{Huibin Chang,\authormark{1,3} Ziqin Rong, \authormark{1} Pablo Enfedaque,\authormark{2} Stefano Marchesini\authormark{2,4} }
%\author{Huibin Chang,\authormark{1,3}  Pablo Enfedaque,\authormark{2} Stefano Marchesini\authormark{2,4} }
\address{\authormark{1} School of Mathematical Sciences, Tianjin Normal University, Tianjin, China\\ \authormark{2}Computational Research Division, Lawrence Berkeley National Laboratory, Berkeley, CA, USA\\ \authormark{3}changhuibin@gmail.com \\ \authormark{4}smarchesini@lbl.gov }
\else
\author[a*]{Huibin Chang}{}
\author[a]{Ziqin Rong}{}
\author[b]{Pablo Enfedaque}{}
\cauthor[b*]{Stefano}{Marchesini}{changhuibin@gmail.com, smarchesini@lbl.gov}

\aff[a]{School of Mathematical Sciences, Tianjin Normal University, Tianjin, \country{China}}
\aff[b]{Computational Research Division, Lawrence Berkeley National Laboratory, Berkeley, CA, \country{USA}}
\fi

%\author{Huibin Chang\thanks{Corresponding author. School of Mathematical Sciences, Tianjin Normal University, Tianjin, 300387,  China, {\tt E-mail:changhuibin@gmail.com}. The author is currently a visiting researcher of the Computational Research Division at Lawrence Berkeley National Laboratory.}
%\and{Pablo Enfedaque}\thanks{Computational Research Division, Lawrence Berkeley National Laboratory, Berkeley, CA 94720, USA,    {\tt Email:penfedaque@lbl.gov}},
%    \and {Stefano Marchesini}\thanks{Corresponding author. Computational Research Division, Lawrence Berkeley National Laboratory, Berkeley, CA 94720, USA,    {\tt Email:smarchesini@lbl.gov}}    %\hskip 4.5cm {\color{red}\today}
%    }

% Set the article type
%\begin{document}
%\maketitle
%\slugger{sisc}{xxxx}{xx}{x}{x--x}%slugger should be set to mms, siap, sicomp, sicon, sidma, sima, simax, sinum, siopt, sisc, or sirev


\begin{abstract}
Spectroscopic ptychography is a powerful technique to determine the chemical composition of a sample with high spatial resolution. In spectro-ptychography, a sample is rastered through a focused x-ray beam with varying photon energy so that a series of phaseless diffraction data are recorded. Each chemical component in the material under investigation has a characteristic absorption and phase contrast as a function of photon energy. Using a dictionary formed by the set of contrast functions of each energy for each chemical component, it is possible to obtain the chemical composition of the material from  high resolution multi-spectral images. This paper presents SPA (Spectroscopic Ptychography with ADMM), a novel algorithm to iteratively solve the spectroscopic blind ptychography problem. We design first a nonlinear spectro-ptychography model based on Poisson maximum likelihood, and construct then the proposed method based on fast iterative splitting operators. SPA can be used to  retrieve spectral contrast when considering both a known or an incomplete (partially known) dictionary of reference spectra. By coupling the redundancy across different spectral measurements, the proposed algorithm can achieve higher reconstruction quality when compared to standard state-of-the-art two-step methods. We demonstrate how SPA can recover accurate chemical maps from Poisson-noised measurements, and also show its enhanced robustness when reconstructing reduced redundancy ptychography data using large scanning stepsizes.
%\end{abstract*}
\end{abstract}

%%%%%%%%%%%%%%%%%%%%%%%%%%  body  %%%%%%%%%%%%%%%%%%%%%%%%%%
\section{Introduction}\label{intro}

\onecolumn

X-ray spectro-microscopy is a powerful technique to study the chemical and morphological structure of a material at high resolution. The contrast of the material under study is recorded as a function of photon energy, and this spectral absorption contrast can later be used to reveal details about its chemical, orbital or magnetic state \cite{stohr2013nexafs,koningsberger1988x}. The idea is that because different chemical components interact differently with the beam at different energies, the composition map of a sample can be solved by using a measured reference of spectra (dictionary). 

Compared to standard lens-based microscopy, x-ray ptychography can provide much finer spatial resolution, while also providing additional phase contrast of the sample~\cite{nellist1995resolution,chapman1996phase,rodenburg2004phase,rodenburg2007hard}. Ptychography is based on retrieving the phase of diffraction data recorded to a numerical aperture that is far larger than what x-ray optics can technically achieve.
In ptychography, the probe (illumination) is almost never completely known, so a joint recovery problem (sample and probe) is typically considered, referred to as blind ptychography. 
Several algorithms to solve both standard and blind ptychography problems have been published in the literature to solve both standard and blind ptychography problems, which also consider a variety of additional experimental challenges~\cite{maiden2009improved,thibault2009probe,thibault2012maximum,wen2012,marchesini2013augmented,horstmeyer2015solving,hesse2015proximal,odstrvcil2018iterative,chang2018Blind}.

As in standard spectro-microscopy, it is also possible to perform spectroscopic ptychography by recording diffraction data at different x-ray photon energies. In recent years, spectro-ptychography has become an increasingly popular chemical analysis technique \cite{beckers2011chemical,maiden2013soft,hoppe2013high,shapiro2014chemical,farmand2017near,shi2016soft}. However, the standard methodology involves independent ptychographic reconstructions for each energy, followed by component analysis, i.e. spectral imaging analysis based on a known reference spectra or multivariate analysis \cite{adams1986spectral,lerotic2004cluster,shapiro2014chemical,yu2018three}. More recently, a low rank constraint \cite{vaswani2017low} for multi-channel samples was proposed together with a gradient descent algorithm with spectral initialization to recover the higher dimension phase retrieval problem {(without component analysis)}. Another work proposed a hierarchical model with Gaussian-Wishart hierarchical prior and developed a variational expectation-maximization  algorithm \cite{liu2019low}. Also, matrix decomposition based low rank prior \cite{chen2018low} has been exploited to reconstruct dynamic, time-varying targets in Fourier ptychographic imaging.

%\subsection{Motivation for end-to-end scheme}

In this paper we propose a novel technique to solve the blind x-ray spectro-ptychography problem based on coupling the diffraction data from each photon energy and iteratively retrieve the chemical map of the sample. The proposed algorithm, referred to as SPA (Spectroscopic Ptychography with ADMM), works with both completely and partially known reference spectra. The method is designed using the Alternating Direction Method of Multipliers (ADMM)~\cite{glowinski1989augmented,chang2018Blind} framework, employing also total variation (TV) regularization~\cite{rudin1992nonlinear} on the chemical map. Compared with the standard two-step methods, the proposed joint reconstruction algorithm can generate much higher quality results without presenting the phase ambiguity problem inherent to two-step methods\footnote{
{Since in the first step for ptychography reconstruction for the two-step method, each  sample contrast map is recovered independently, that causes to different phase factors for different  maps.}}. The simulation analysis shows the efficient convergence ratio of SPA and demonstrates the increased robustness of the method to large stepsizes, being able to retrieve features lost when using standard two-step methods. The algorithm is described and analyzed with and without TV regularization for both partially and completely known dictionary cases.
 
%    \item Two-step way could fail due to the strong noise of measured frames and weak contrast of the sample for some frequency X-rays.
%    \item Joint illumination so as to allow for large scan stepsize and be robust to noise.
%    \item For ``flat''-sample, end-to-end way can speed up the convergence speed.
%\end{itemize}

\section{Spectroscopic ptychography model}
%\subsection{SP with weak phase approximation}
\onecolumn

Given  $L$ different energies of x-rays going through a sample illuminated by a probe $\omega\in\mathbb C^{\bar m}$, 
a collection of phaseless intensities $\{I_l\}_{l=0}^{L-1}$ are measured in the far field, 
such that with {Poisson fluctuation caused by photon counting,} we have 
%\[
%{I_l}=|\mathcal A(\omega,\mathcal T_l Y_l)|^2~\forall~0\leq l\leq L-1,
%\]
\[
{I_l}=\mathrm{Poi}(|\mathcal A(\omega, Y_l)|^2)~\mathrm{for}~l=0,1,\cdots, L-1,
\]
where  $Y=(Y_0,Y_1,\ldots, Y_{L-1})\in \mathbb C^{N, L}$ is the sample contrast map for each X-ray energy, {\color{black} $\mathcal{A}(w,\cdot)$ is the forward operator for ptychography for a given probe $w$},
$\mathrm{Poi}$ denotes the Poisson-noise contamination, {\color{black} and the notations $|\cdot|, (\cdot)^2$ denoting the pointwise absolute and square values of the vector respectively. Note that the probe $\omega$, and each column of contrast maps $\{Y_l\}$ are all 2D images, written as  vectors by a lexicographical order. The relationship between the contrast map $Y_{l}$ observed by ptychography at each energy and an unknown sample elemental map $X$ made of $C$ elements is governed by the  spectral contrast of each element, stored in a ``dictionary'' $D$ of known tabulated values.

Specifically, following similar notation as in \cite{chang2018partially}, 
the bilinear operator  $\mathcal A:\mathbb C^{\bar m}\times \mathbb C^{N}\rightarrow \mathbb C^{m}$ 
%and
 %$\mathcal A_j:\mathbb C^{\bar m}\times \mathbb C^{N}\rightarrow \mathbb C^{\bar m}~\forall 0\leq j\leq J-1$,  
 is defined as
 \[
 \mathcal A(\omega,u):=\Big((\mathcal F(\omega\circ \mathcal S_0 u))^T, (\mathcal F(\omega\circ \mathcal S_1 u))^T,\cdots, (\mathcal F(\omega\circ \mathcal S_{J-1} u))^T\Big)^T,\]
 }
 %$\mathcal A_j(\omega,u):=\mathcal F(\omega\circ \mathcal S_j u)$, 
 {where $\mathcal F$ denotes the discrete Fourier transform, $\circ$ denotes the Hadamard product (pointwise multiplication) of two vectors, and $\mathcal S_j\in \mathbb R^{\bar m, m}$ is a binary matrix that defines a small window with the index $j$ and size $\bar m$ over the entire image $u$ (taking small patches out of the entire image).}
%, and $\mathcal T_l$ denotes the translations.
%{\color{red}Here the translations of the illumination could produce more redundancy of scan, and meanwhile it helps to reduce the grid pathology. [Proof]} 
%Generalizations with a known transformation when changing energy is omitted for simplicity.

For different energies, assuming that a spectrum dictionary $D\in\mathbb C^{C, L}$ (or its absorption part) is measured in advance, having $C$ components for different materials or particles, and given a  sufficiently thin specimen, the sample contrast maps can be approximated  by first-order Taylor expansion $\exp(XD)\approx \bm 1+XD $ \footnote{$\bm 1$ denotes the matrix with all elements being one, of the same size of $Y$. $\exp(\cdot)$ denotes the pointwise exponentiation of the matrix.} as 
\begin{equation}
Y=\bm 1+XD,
\label{eqDecomp}
\end{equation}
{\color{black}with $
X=(X_0,X_1,\ldots, X_{C-1})\in\mathbb R_+^{N,  C}$ being the elemental thickness map of the sample} {(each column of thickness map denotes the thickness of each components in the object)}.
%where $X$ is the thickness map for the materials.

%\subsection{Model}

%We also assume that there exists an known empty region (e.g. vacuum), i.e. $X|_{\Omega}=0$, such that the phase factors can be removed for different energies.
To determine the thickness map $X$, with a completely known spectrum $D$, one has to solve the following problem  : 
\begin{equation}
\text{To find $X$ and $\omega$, ~~s.t.~} {|\mathcal A(\omega, Y_l)|^2\approx I_l,}~~ Y=\bm 1+XD,~~X\in\mathscr X,
\end{equation}
 with non-negative thickness constraint set {\color{black}$\mathscr X=\{X=(X_{n,c})\in \mathbb R^{N, C}: X_{n,c}\geq 0,\ 0\leq n\leq N-1, 0\leq c\leq C-1\}.$}
 Letting  the illumination  be normalized, i.e. $\omega\in\mathscr W:=\{\omega\in\mathbb C^{\bar m}:~\|\omega\|=1\}$, 
the total variation  regularized nonlinear optimization model can be established by assuming the piecewise smoothness of the thickness map as: 
\begin{equation}\label{SP}
\begin{split}
\mathrm{SP:}&\min_{\omega,X,Y}\delta \sumMy\nolimits_c \mathrm{TV}(X_c)+\sumMy\nolimits_l\mathcal G(\mathcal A(\omega, Y_l);I_l)+\mathbb I_{\mathscr X}(X)+\mathbb I_{\mathscr W}(\omega),~~\text{s.t.}~~Y=\bm 1+X D,
\end{split}
\end{equation}
where 
%$\mathcal G(z;f):=\tfrac12\||z|-\sqrt{f}\|^2$ 
{\color{black}$\mathcal G(z;f):=\tfrac12\sum_{n=0}^{N-1}  \big(|z_n|^2-f_n\log(|z_n|^2)\big)~\forall z=(z_0, z_1, \cdots, z_{N-1})^T\in\mathbb C^N,$ $f=(f_0,f_1,\cdots,f_{N-1})^T\in\mathbb R^N,$} derived from the maximum likelihood estimate of Poisson noised data \cite{chang2016Total},  { $\mathrm{TV}$ denotes the standard total variation semi-norm \cite{rudin1992nonlinear} to enforce the piecewise smooth structure of $X_c$ (the $c^{th}$ column of the mixing matrix (thickness map) denoted in \eqref{eqDecomp}), and $\delta$ is a positive constant to balance the regularization and fitting terms (Bigger $\delta$ produces stronger smoothness).  Here $\mathbb I_{\mathscr X}(X),\mathbb I_{\mathscr W}(\omega)$ denotes the indicator functions, with
$
\mathbb I_{\mathscr X}(X)=0~\text{if~} X\in\mathscr X; \mathbb I_{\mathscr X}(X)=+\infty,~\text{~otherwise.} 
$
We remark that it is a convenient way to enforce hard constraints within an optimization formulation.
}
 
Experimentally,  only the real values part (absorption)  of the dictionary $D_r$ are measured.
As $X$ is real-valued, we consider the following relation:
\[
\Re(Y)=D_r X+\bm 1,%,~~\Im(Y)=D_i X,
\]
where $D_r:=\Re(D)$. %and $D_i=\Im(D).$
Similarly, we derive the following spectroscopic ptychography  with incomplete dictionary (SPi) as 
\begin{equation}\label{eq:spi}
\begin{split}
\mathrm{SPi:}&\min_{\omega,X,Y}\delta \sumMy\nolimits_c \mathrm{TV}(X_c)+\sumMy\nolimits_l\mathcal G(\mathcal A(\omega, Y_l);I_l)+\mathbb I_{\mathscr X}(X)+\mathbb I_{\mathscr W}(\omega),~\text{s.t.}~~\Re(Y)=\bm 1+D_r X.
\end{split}
\end{equation}

\begin{rem}
Unlike solving the ptychography imaging independently for each energy, we use the low-rank structure of the recovery results of different energies, i.e. the rank of the matrix $Y-\bm 1$ is no greater than that of $X$.   
\end{rem}

\vskip .2in
{\color{black}
\begin{table}[]
    \caption{Main variables and operators defined in section 2}
    \begin{center}
          %  \scalebox[.7]{
    \begin{tabular}{|l|l|}
    \hline\\
    Notations& Explanations \\
         \hline \hline
           $\{I_l\}_{l=0}^{L-1}$& Measured intensities\\
    \hline
     $X\in\mathbb R_+^{N, C}$& 
     \makecell[l]{Elemental thickness maps of the sample} \\
     %\\
     %(Each column denotes thickness of each components in the object)}\\
        \hline
   $D\in\mathbb C^{C,L}$& Spectrum dictionary \\
   \hline 
   $Y\in \mathbb C^{N, L}$& Sample spectral contrast maps \\     
   \hline 
     $\omega\in\mathbb C^{\bar m}$& Probe\\ 
                \hline
   $\mathcal S_j\in \mathbb R^{\bar m, m}$ & Binary matrix to take image patches\\
        \hline
        $\mathcal A:\mathbb C^{\bar m}\times \mathbb C^{N}\rightarrow \mathbb C^{m}$  & Forward operator for ptychography\\
        \hline
        ${\mathcal{G}}(\mathcal{A}(w,u), I_{l})$ & Poisson likelihood estimation
        \\
     \hline
     $D_r\in\mathbb R^{C,L}$& Real part of spectrum dictionary 
     \\
   %$\mathcal A(\omega,u):=(\mathcal A_0^T (\omega,u), \mathcal A_1^T(\omega,u),\cdots, \mathcal A_{J-1}^T(\omega,u))^T,$ \\
% $\mathcal A_j(\omega,u):=\mathcal F(\omega\circ \mathcal S_j u)$
   \hline
   $\mathscr X$& Non-negative thickness constraint\\
   \hline
   $\mathscr W$& Normalized constraint of the probe\\
   \hline
   TV & Total variation regularization\\
   \hline
  \end{tabular}
  %}
    \end{center}
        \label{tab:my_label}
\end{table}
}
%\subsection{Dictionary based ptychography}

%\subsection{Low-rank regularized ptychography}
%\[
%\mathrm{LRHSP:}~\min_{\omega,Y}\sum\nolimits_l\mathcal G(Z_l;I_l)+\eta\|Y-\bm %1\|_*,~~s.t. Z_l=\mathcal A(\omega,Y_l),
%\]
%with the nuclear norm $\|\cdot\|_*$ enforcing the low rank property.
%\section{Iterative algorithm}

\section{Proposed iterative algorithm}
\onecolumn

%\subsection{Review of ADMM and BP-ADMM}
ADMM \cite{glowinski1989augmented} is a powerful and flexible tool  that has already been applied to both ptychography  \cite{wen2012,chang2018Blind} and phase tomography problems \cite{chang2019iterative,aslan2019joint}. 
In this work we also adopt the ADMM framework to design an iterative joint spectro-ptychography solution. We construct the proposed algorithm considering both complete and incomplete dictionary cases.

\subsection{Complete dictionary}
Based on the spectro-ptychography model~\eqref{SP} for a complete dictionary of spectra, we design the proposed SPA algorithm (Spectroscopic Ptychography with ADMM) as described below. 

Let $DD^*\in\mathbb C^{C, C}$ be non-singular, {where $D^*$ denotes the Hermitian
matrix of $D$, i.e. $D^*:=\mathrm{conj}(D^T)$.}
Considering the constraint in \eqref{eqDecomp}, the following equivalent form can be derived
\begin{equation}\label{eq:0}
X= (Y-\bm 1) \hat D,
%X=Y\hat D,
\end{equation}
with $\hat D:=D^*(DD^*)^{-1}\in\mathbb C^{L, C}$. Accordingly, the following equivalent model can be considered, by introducing  auxiliary variables $\{Z_l\}$:
\begin{equation}
\begin{split}
&\min_{\omega,X,Y}\delta \sumMy\nolimits_c \mathrm{TV}(X_c)+\sumMy\nolimits_l\mathcal G(Z_l;I_l)+\mathbb I_{\mathscr X}(X)+\mathbb I_{\mathscr W}(\omega),\\
&~~\text{s.t.}~~Z_l=\mathcal A(\omega, Y_l), X=(Y-\bm 1)\hat D~\forall 0\leq l\leq L-1.
\end{split}
\label{eq:1}
\end{equation}
%where 
%$\mathcal G(z;f):=\tfrac12\||z|-\sqrt{f}\|^2$ 
%$\mathcal G(z,f):=\tfrac12\langle \bm 1,  |z|^2-f\circ\log(|z|^2)\rangle$.
%$\mathcal P$ denotes the restriction operator (binary matrix), such that the boundary layer is neglected.
The benefit of considering \eqref{eq:0} instead of \eqref{eqDecomp} lies in the fact that: (i) the multiplier will be a low-dimensional variable, since the dimension of $Y$ is much higher than that of $X$, and (ii) the subproblem w.r.t. the variable $X$ can be more easily solved.

An equivalent saddle point problem for \eqref{eq:1}, based on the augmented Lagrangian, can be derived as:  
\[
\begin{split}
&\max_{\Lambda,\Gamma}\min_{\omega,X,Y,Z}\mathscr L_{\lambda,\beta}(\omega,X,Y, Z, \Lambda,\Gamma)\\
:=&\delta \sumMy\nolimits_c \mathrm{TV}(X_c)+\sumMy\nolimits_l\mathcal G(Z_l;I_l)+\mathbb I_{\mathscr X}(X)+\mathbb I_{\mathscr W}(\omega),\\
&+\sumMy\nolimits_l\left(\lambda\Re\langle Z_l-\mathcal A(\omega, Y_l), \Lambda_l\rangle+\tfrac{\lambda}{2}\|Z_l-\mathcal A(\omega, Y_l)\|^2\right),\\
&+\beta\Re\langle  X-(Y-\bm 1)\hat D, \Gamma\rangle+\tfrac{\beta}{2}\|X-(Y-\bm 1)\hat D\|^2,
\end{split}
\]
with the multipliers $\Lambda:=(\Lambda_0,\cdots,\Lambda_{L-1})$ and $\Gamma$, where $\langle \cdot,\cdot\rangle$ denotes the inner product of vectors and matrices (trace norms) respectively.

The above saddle point problem can be solved by alternating minimization and update of the multipliers. 
We first define each sub-minimization problem. The $\omega-$subproblem, with the additional proximal term, can be expressed as 
\[
\begin{split}
&\omega^\star:=\arg\min\mathscr L_{\lambda,\beta} (\omega, X,Y,Z,\Lambda,\Gamma)\\
&=\arg\min\nolimits_{\omega} \tfrac12\sumMy\nolimits_{l}\|Z_l+\Lambda_l-\mathcal A(\omega,  Y_l)\|^2+\mathbb I_{\mathscr W}(\omega)\\
&=\arg\min\nolimits_{\omega\in \mathscr W}  \tfrac12\sumMy\nolimits_{l,j}\|\mathcal F^*(Z_{l,j}+\Lambda_{l,j})-\omega\circ\mathcal S_j  Y_l\|^2.\\
%&=\arg\min\nolimits_{\omega}  \tfrac12\sum\nolimits_{l,j}\|\mathcal T_l^T\mathcal F^*(Z_{l,j}+\Lambda_{l,j})-\omega\circ\mathcal T_l^T\mathcal S_j Y_l\|^2+\tfrac{\gamma_1}{2}\|\omega-\omega_0\|^2.
\end{split}
\]
The first-order gradient of the above least squares problem (without constraint) is given as
\[
{H(\omega):=}\mathrm{diag}(\sumMy\nolimits_{l,j}|\mathcal S_j  Y_l|^2) \omega-\sumMy\nolimits_{l,j} \mathcal F^*(Z_{l,j}+\Lambda_{l,j})\circ  \mathcal S_j  Y_l^*.
\]
Consequently, the projected gradient descent scheme with preconditioning can be derived as 
\begin{equation}
\begin{split}
{ \omega_{s+1}=\mathrm{Proj}_{\mathscr W}\left(\omega_s-\frac{H(\omega_s)}{(\sumMy\nolimits_{l,j}|\mathcal S_j Y_l|^2) +\gamma_1 \mathbf 1}\right)}~~s=0,1, \cdots,
%&=\mathrm{Proj}_{\mathscr W}\left(\tfrac{\gamma_1\omega_s}{(\sumMy\nolimits_{l,j}|\mathcal S_j Y_l|^2) +\gamma_1 \mathbf 1}+\tfrac{\sumMy\nolimits_{l,j} \mathcal F^*(Z_{l,j}+\Lambda_{l,j})\circ  \mathcal S_j  Y_l^*}{(\sumMy\nolimits_{l,j}|\mathcal S_j Y_l|^2) +\gamma_1 \mathbf 1}\right)~~s=0,1, \cdots,
\end{split}
\end{equation}
with parameter $\gamma_1>0$ in order to get rid of division by zeros, and $\mathrm{Proj}_{\mathscr W}(\omega):=\frac{\omega}{\|\omega\|}$.  Here the parameter $\gamma_1$ heuristically set to be a small scalar related with the maximum value of the $\sumMy\nolimits_{l,j}|\mathcal S_j Y_l|^2 $, e.g. {\color{black}$\gamma_1=0.1 \times\|\sum\nolimits_{l,j}|\mathcal S_j Y^k_l|^2\|_\infty.$}

The $X-$subproblem can be expressed as
\[
\begin{split}
X^\star&:=\arg\min\nolimits_X \mathscr L_{\lambda,\beta}(\omega, X, Y, Z, \Lambda, \Gamma) \\
&=\arg\min\nolimits_{X}
 \sumMy\nolimits_c \left(\tfrac{\delta}{\beta}\mathrm{TV}(X_c)+\tfrac12\|X_c-\Re((Y-\bm 1)\hat D-\Gamma)_c\|^2\right)+\mathbb I_{\mathscr X}(X),\\
\end{split}
\]
{where $(\cdot)_c$ denotes the $c^{th}$ column of a matrix.}
Since it is common practise to solve the total variation denoising problem by using a first-order operator-splitting algorithm \cite{wu2011augmented,chambolle2011first}, we directly give the approximate solution below:
\[
X_c^\star=\max\{0, \mathrm{Denoise}_{\delta/\beta}(\Re(((Y-\bm 1)\hat D-\Gamma))_c)\}~\forall 0\leq c\leq C-1,
\]
with
$\mathrm{Denoise}_{\nu}(u_0):=\arg\min_u \nu\mathrm{TV}(u)+\tfrac12\|u-u_0\|^2.$
{  Here we remark that to seek for exact solution with this positivity constraint, one may need more auxiliary variables and inner loops as \cite{Chan2013}. For simplicity, we did not exactly solve the constraint problem, and instead, the above approximation is derived by the standard TV-L2 denoising without constraint and then a projection to the positivity constraint set.  
}

The $Y-$subproblem, with additional proximal term $\tfrac{\gamma_2}{2}\|Y-Y_0\|^2$ and previous iterative solution $Y_0$, is expressed as
\[
\begin{split}
&Y^\star:=\arg\min_Y \mathscr L_{\lambda,\beta} (\omega, X, Y, Z, \Lambda, \Gamma)\\
&=\arg\min\nolimits_{Y}\tfrac{\lambda}{2}\sumMy\nolimits_l\|\mathcal A(\omega,Y_l)-(\Lambda_l+Z_l)\|^2+\tfrac{\beta}{2}\| Y\hat D-(\Gamma+X+\bm 1\hat D)\|^2+\tfrac{\gamma_2}{2}\|Y-Y_0\|^2\\
&=\arg\min\nolimits_{Y}\tfrac{\lambda}{2}\sumMy\nolimits_l\|\omega\circ \mathcal S_j Y_l-\mathcal F^*(\Lambda_l+Z_l)\|+\tfrac{\beta}{2}\| Y\hat D-(\Gamma+X+\bm 1\hat D)\|^2+\tfrac{\gamma_2}{2}\|Y-Y_0\|^2\\
&=\arg\min\nolimits_{Y}\tfrac{\lambda}{2}\sumMy\nolimits_l\| \mathcal S_j^T\omega\circ  Y_l-\mathcal S_j^T\mathcal F^*(\Lambda_l+Z_l)\|+\tfrac{\beta}{2}\| Y\hat D-(\Gamma+X+\bm 1\hat D))\|^2+\tfrac{\gamma_2}{2}\|Y-Y_0\|^2,\\
\end{split}
\]
where $\gamma_2$ is a positive scalar similarly to the parameter $\gamma_1$. 

%We consider a special case that all the channels share same scan geometry, i.e. $\mathcal T_l$ is identity operator, 
By calculating the first-order gradient of the above least squares problem,  one has 
\[
\begin{split}
\mathrm{diag}(\lambda\sumMy\nolimits_j |\mathcal S^T_j \omega|^2+\gamma_2 \mathbf I)Y+\beta  Y \hat D\hat D^*=\lambda Q+\gamma_2 Y_0+\beta(\Gamma+X+\bm 1\hat D)\hat D^*,
\end{split}
\]
{with identity operator $\mathbf I,$}
where $ Q:=(Q_0, Q_1,\ldots,Q_{L-1})\in\mathbb C^{N, L}$ with $Q_l:=\sumMy\nolimits_j\mathcal S_j^T (\omega^*\circ \mathcal F^*(\Lambda_l+Z_l)),$ which  is actually the Sylvester equation \cite{Sylvester,Simoncini2016}. 

Assuming that the positive Hermitian $\hat D\hat D^*$ has the singular-value-decomposition (SVD) as $\hat D\hat D^*=V\mathscr S V^*,$ with diagonal matrix (diagonal elements are singular values) $\mathscr S\in \mathbb R^{L, L}$ and unitary matrix $V\in \mathbb C^{L, L}$, by introducing $\hat Y:=YV,$ we derive:
\[
\begin{split}
\mathrm{diag}(\lambda\sumMy\nolimits_j |\mathcal S^T_j \omega|^2+\gamma_2 \mathbf I)\hat Y+\beta  \hat Y\mathscr S=(\lambda Q+\gamma_2 Y_0+\beta(\Gamma+X+\bm 1\hat D)\hat D^*)V,
\end{split}
\]
such that the closed form solution can be expressed as
\begin{equation}\label{slv1}
Y^\star=\hat Y^\star V^*, 
\end{equation}
where $
\hat Y^\star:=(\hat Y^\star_0,\cdots,\hat Y^\star_{L-1})\in\mathbb C^{N, L},$ and
\begin{equation}
\label{slv2}
\hat Y^\star_l=\dfrac{((\lambda Q+\gamma_2 Y_0+\beta(\Gamma+X+\bm 1\hat D)\hat D^*)V)_l}{(\lambda\sumMy\nolimits_j |\mathcal S^T_j \omega|^2+\gamma_2 \mathbf 1)+\beta\mathscr S_{l,l}\bm 1}~~\forall~0\leq l\leq L-1.
\end{equation}

\begin{comment}
If the scan geometries are different for different channels, we have 
\[
\begin{split}
B_\omega Y+\beta  Y \hat D\hat D^*=\lambda \bar Q+\gamma_2 Y_0+\beta(\Gamma+X+\bm 1\hat D)\hat D^*.
\end{split}
\]
where
$B_\omega Y:=[W_0 Y_0, W_1Y_1, \cdots, W_{L-1}Y_{L-1}]$ and
$W_l:=\mathrm{diag}(\mathcal T_l^T\sumMy\nolimits_j |\mathcal S^T_j \omega|^2)+\tfrac{\gamma_2}{\lambda} \mathbb I$,
and $ \bar Q:=(\bar Q_0, \bar Q_1,\ldots,\bar Q_{L-1})\in\mathbb C^{N, L}$ with $\bar Q_l:=\sumMy\nolimits_j\mathcal T_l^T \mathcal S_j^T ( \omega^*\circ \mathcal F^*(\Lambda_l+Z_l)).$ We proposed an preconditioned gradient descent method below:
\begin{equation}
\begin{split}
Y&\leftarrow Y- \tau B_\omega^{-1}G Y\\
%&\leftarrow (1-\tau) Y-\tau \beta B_\omega^{-1}  Y \hat D\hat D^*+\tau B_\omega^{-1}(\lambda \bar Q+\gamma_2 Y_0+\beta(\Gamma+X+\bm 1\hat D)\hat D^*),  
\end{split}
\end{equation}
with stepsize $\tau$, 
 $G Y:=AY-(\lambda \bar Q+\gamma_2 Y_0+\beta(\Gamma+X+\bm 1\hat D)\hat D^*),$
and $A Y:=B_\omega Y+\beta  Y \hat D\hat D^*.$

The stepsize can be selected as Cauchy stepsize as
\[
\tau:=\dfrac{\Re\langle GY, B_\omega GY \rangle}{\Re\langle AB_\omega GY, B_\omega GY \rangle}
\]
\end{comment}

%where the preconditiner $P:=\mathcal B_\omega^{-1}$

%The above system is positive such that the solution is unique. 

For the $Z-$subproblem, we have \cite{chang2016Total}:
\[
Z^\star:=\arg\min_{Z} \sumMy\nolimits_l\mathcal G(Z_l;I_l)+\tfrac{\lambda}{2}\sumMy\nolimits_l\|Z_l-(\mathcal A(\omega,Y_l)-\Lambda_l)\|^2,
\]
which gives
\[
%Z_l^\star=\tfrac{\sqrt{I_l}+\lambda|\hat Z_l|}{1+\lambda}\circ\mathrm{sign}(\hat Z_l),
Z_l^\star=\tfrac{\sqrt{4(1+\lambda)I_l+\lambda^2|\hat Z_l|^2}+\lambda|\hat Z_l|}{2(1+\lambda)}\circ\mathrm{sign}(\hat Z_l),
\]
with $\hat Z_l:=\mathcal A(\omega,Y_l)-\Lambda_l$.
Based on the above calculations and the update of multipliers, the baseline SPA algorithm is summarized in the appendix.

\subsection{Incomplete dictionary }
\onecolumn

{  A complete dictionary is often difficult to obtain without an independent experiment prior to a spectro-ptychography experiment. 
The materials components and their chemical states are often not known in advance. Moreover  the  real part of the refractive index  component  is  often  not  well  known  \cite{henke1993x}.  It  is  more difficult to measure because it requires an interferometric or reflectometry measurements rather than simple absorption spectroscopy measurements and reflectometry experiments are less commonly done. While the Cramers-Kronig  relationships relate  real and imaginary parts,  the relationship requires an spectral measurement from 0 to infinity which is not possible to measure in finite time;  Standard techniques to extend absorption spectra can only produce approximate values in the imaginary component. Hence, it is attractive in practice to provide a version working with the real part only.
}

In this subsection, we propose a variation of the SPA algorithm to solve the joint spectro-ptychography problem when the dictionary of spectra is only partially know, based on the model proposed in~\eqref{eq:spi}.
By assuming that $D_r$ has full row-rank, i.e. $D_rD_r^*$ is non-singular, with known $\hat D_r:=D_r^T(D_rD_r^T)^{-1}$ in advance, we have:
\[
X=\Re(Y-\bm 1)\hat D_r.
\]
Consequently, the following equivalent problem can be solved instead of \eqref{eq:spi}:
\begin{equation}
\begin{split}
\min_{\omega,X,Y}\delta \sumMy\nolimits_c \mathrm{TV}(X_c)+\sumMy\nolimits_l\mathcal G(\mathcal A(\omega, Y_l);I_l)+\mathbb I_{\mathscr X}(X)+\mathbb I_{\mathscr W}(\omega),~\text{s.t.}~~X=\Re(Y-\bm 1)\hat D_r.
\end{split}
\label{eq:SPiVar}
\end{equation}

Similarly to the previous subsection,  introducing the multiplier $\Gamma_r$  and auxiliary variable $Z$  yields the saddle point problem below, with the help of the augmented Lagrangian of \eqref{eq:SPiVar}: 
\[
\begin{split}
&\max_{\Lambda,\Gamma_r}\min_{\omega,X,Y,Z}{\mathscr {\widetilde L} }_{\lambda,\beta}(\omega,X,Y, Z, \Lambda,\Gamma_r):=\delta \sumMy\nolimits_c \mathrm{TV}(X_c)+\sum\nolimits_l\mathcal G(Z_l;I_l)+\mathbb I_{\mathscr X}(X)+\mathbb I_{\mathscr W}(\omega)\\
&\qquad\qquad+\sum\nolimits_l\left(\lambda\Re\langle Z_l-\mathcal A(\omega,Y_l), \Lambda_l\rangle+\tfrac{\lambda}{2}\|Z_l-\mathcal A(\omega,Y_l)\|^2\right)\\
&\qquad\qquad+\beta\langle X-\Re(Y-\bm 1)\hat D_r, \Gamma_r \rangle+\tfrac{\beta}{2}\|X-\Re(Y-\bm 1)\hat D_r\|^2.\\
%&\qquad\qquad+\beta\langle X-\Im(Y)\hat D_i, \Gamma_i \rangle+\tfrac{\beta}{2}\|X-\Im(Y)\hat D_i\|^2,\\
\end{split}
\]

Below, we focus only on the differences with respect to the Algortihm 1. 
For the $X-$subproblem, we have 
\[
\begin{split}
&X^\star:=\arg\min  \tfrac{\delta}{\beta} \sumMy\nolimits_c \mathrm{TV}(X_c)+ \mathbb I_{\mathscr X}(X)+\tfrac{1}{2}\|X-(\Re(Y-\bm 1)\hat D_r-\Gamma_r)\|^2.
\end{split}
\]
Hence we get 
\[
X_c^\star=\max\{0, \mathrm{Denoise}_{\delta/\beta}((\Re(Y-\bm 1)\hat D_r-\Gamma_r)_c)\}~\forall 0\leq c\leq C-1.
\]

For the $Y-$subproblem with proximal terms $\|Y-Y_0\|^2$ , we have
\[
\begin{split}
&Y^\star:=\arg\min\nolimits_{Y}\tfrac{\lambda}{2}\sum_{j,l}\|\mathcal S_j^T\omega\circ  Y_l-\mathcal S_j^T\mathcal F^*(\Lambda_l+Z_l)\|^2+\tfrac{\beta}{2}\|X-\Re(Y-\bm 1)\hat D_r+\Gamma_r\|^2+\tfrac{\hat \gamma_2}{2}\|Y-Y_0\|^2,\\
\end{split}
\]
which results in the following equations w.r.t. the real and imaginary parts, respectively:
\begin{equation*}%\left\{ 
\begin{cases}
&\mathrm{diag}(\lambda\sum\nolimits_j |\mathcal S^T_j \omega|^2+\hat\gamma_2 \mathbf 1)\Re(Y)+\beta  \Re(Y) \hat D_r\hat D_r^T=\lambda \Re(Q)+\hat\gamma_2 \Re(Y_0)+\beta(\Gamma_r+X+\bm 1\hat D_r)\hat D_r^T    \\
&\mathrm{diag}(\lambda\sum\nolimits_j |\mathcal S^T_j \omega|^2+\hat\gamma_2 \mathbf 1)\Im(Y)=\lambda \Im(Q)+\hat\gamma_2 \Im(Y_0).%\end{align*}\right.
\end{cases}
\end{equation*}
%\]
Then, the real part of $Y$ can be solved by \eqref{slv1}-\eqref{slv2}, while the imaginary part can be simply computed by
\[
\Im(Y)=\frac{\lambda \Im(Q)+\hat\gamma_2 \Im(Y_0)}{\lambda\sum\nolimits_j |\mathcal S^T_j \omega|^2+\hat\gamma_2 \mathbf 1}.
\]

The overall SPA algorithm with an incomplete dictionary is summarized in the appendix.
%Algorithm 2.

%\section{Extensions}

\section{{\color{black} Simulation and Reconstruction Results}}
\onecolumn

\begin{figure}
\begin{center}
\begin{tabular}{cccc}%\centering
\includegraphics[width=.07\textwidth]{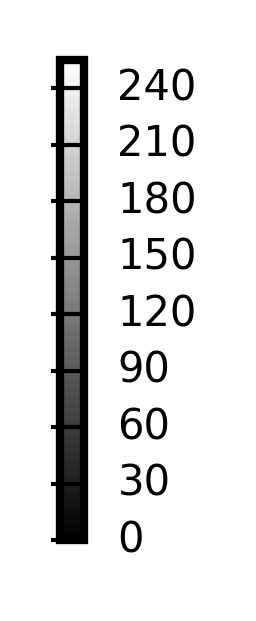}&
\includegraphics[width=.25\textwidth]{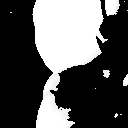}& 
\includegraphics[width=.25\textwidth]{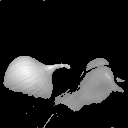}&
\includegraphics[width=.25\textwidth]{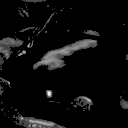}\\
&(a)&(b)&(c)
\end{tabular}
\end{center}
\caption{(a)-(c) Truth for the three different materials.
}\label{fig0}
\end{figure}

%Try (1) more spectral, and (2) low contrast images!
In the simulation analysis of the proposed algorithms we consider the synthetic thickness maps of three different materials, extracted from three (RGB) channels of a natural color image (after thresholding and shift, consisting of $256\times 256$ pixels), shown in Fig. \ref{fig0}. 
The real part of the spectrum dictionary (for two different materials, PMMA {(polymethyl methacrylate)} and PS {(polystyrene)},  plus a constant w.r.t. ten different energies) was measured at the Advanced Light Source \cite{yan2013accurate}, and the imaginary part was derived using 
Kramers-Kronig relations \cite{kronig1926theory}. Both real and imaginary part dictionaries are shown in Fig. \ref{fig1}. 

\begin{figure}
\begin{center}
\begin{tabular}{ccc}%\centering
\includegraphics[width=.25\textwidth]{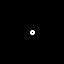}&
\includegraphics[width=.07\textwidth]{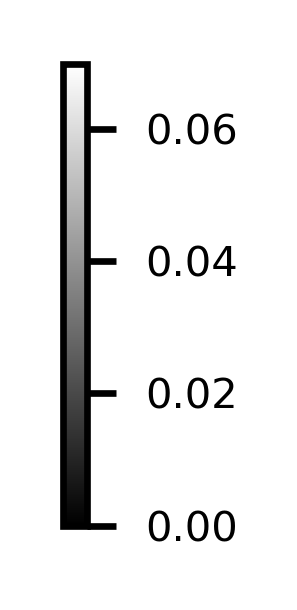}&
\includegraphics[width=.25\textwidth]{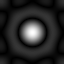}\\
(a)&&(b)
\end{tabular}
\end{center}
\caption{(a) Lens (binary), (b) probe ($64\times 64$ pixles).
}\label{fig0-0}
\end{figure}

\begin{figure}
\begin{center}
\begin{tabular}{cc}%\centering
\includegraphics[width=.45\textwidth]{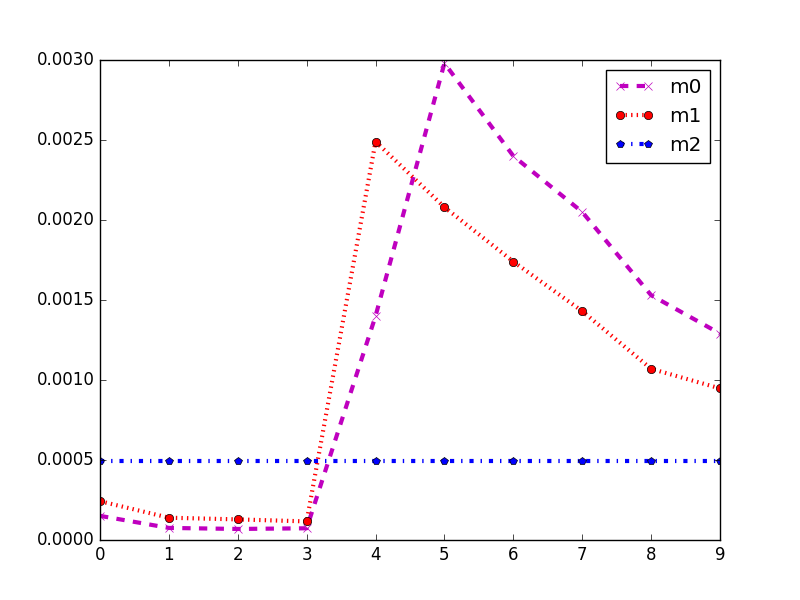} &
\includegraphics[width=.45\textwidth]{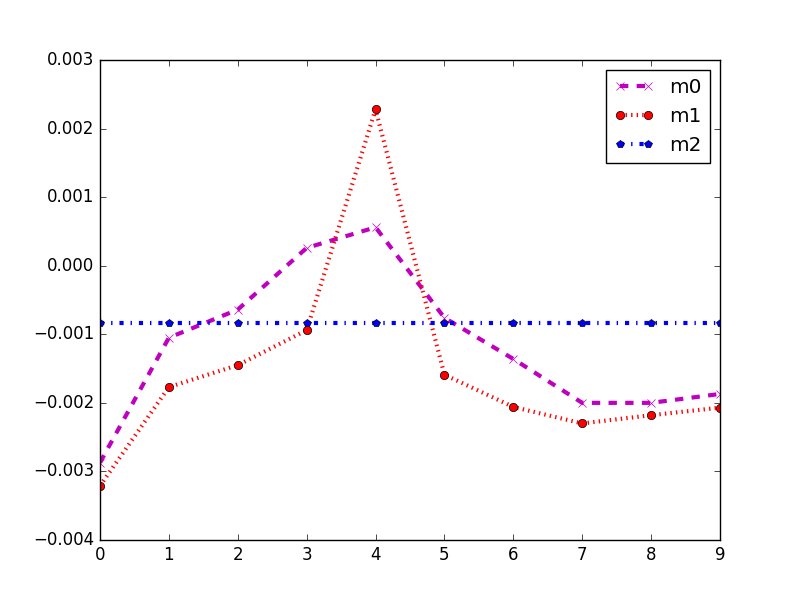}\\
(a)&(b)
\end{tabular}
\end{center}
\caption{Spectrum dictionaries (three different materials m0 (PMMA), m1 (PS) and m2 (constant)), with real part (a) and imaginary part (b). x- and y-axis denote spectrum and different energies, respectively.}
\label{fig1}
\end{figure}

The ptychography measurements are simulated with Poisson noise contamination, using a single grid scan at each energy. 
{ 
{A standard zone-plate} with annular shape diffracts an image (Fig. \ref{fig0-0} ) onto the detector after going through an order-sorting aperture. The zone plate annular aperture is mapped onto the detector by geometric magnification as \textit{outer-diameter} (in microns) 
corresponds to an annular ring on the detector of dimension \textit{outer-diameter}/ detector-pixel-size $\times$ detector-distance / focal-distance . The illumination probe (Fig. \ref{fig0-0}b)  has beam width (full width-half max, FWHM) of 16 pixels.  The relationship between pixels and actual dimensions in the Far-Field approximation is as follows: 
illumination pixel (real-space) dimensions =
 (wavelenght $\times $ 
 detector-distance)/ detector-number-of-pixels $\times$ detector-pixel-size. The zone plate distance to the sample is assumed to be adjusted proportionally with energy to keep the sample in focus as usually done experimentally. We also assume that the detector distance is adjusted to maintain the spatial frequencies on the same detector pixels.
 }

In order to evaluate the recovered results, the signal-to-noise ratio (SNR) in dB is used, which is denoted below:
 \begin{equation}
 \mathrm{SNR}(X, X_g)=-10\log_{10}{\| X-X_g\|^2}/{\|X\|^2},
 \end{equation}
where $X_g$ corresponds to the ground truth thickness.

We compare the proposed iterative SPA algorithm with the standard two-step method. The two-step method consist of (i) perform ptychography reconstruction using a joint illumination, then (ii) perform spectroscopy analysis with a known dictionary (or known real part), and finally (iii) correct the phase ambiguity for different energies. 

When assessing the performance of SPA, we consider both with and without regularization cases, where we simply set the regularization parameter $\delta=0$ and slightly adjust the  algorithm by  replacing Step 2 with  
\begin{equation}
X^{k+1}=\max\{0, \Re((Y^k-\bm 1)\hat D-\Gamma^k)\},    
\end{equation} for baseline SPA and 
\begin{equation}
X^{k+1}=\max\{ 0, \Re(Y^k-\bm 1)\hat D_r-\Gamma_r^k \},    
\end{equation}
for the incomplete dictionary case.

\subsection{Reconstruction quality}
\onecolumn

\begin{figure}
\begin{center}
\begin{tabular}{cccc}
{\begin{overpic}[width=.25\textwidth]{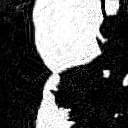}
\put(8,25) {\tikz\draw[red,thick,dashed] (0,0) circle (.3);}
\put(35,40){\tikz\draw[blue,thick,dashed] (0,0) circle(.3);}
\end{overpic}}&
{\begin{overpic}[width=.25\textwidth]{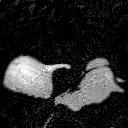}
\put(8,25) {\tikz\draw[red,thick,dashed] (0,0) circle (.3);}
\put(55,15){\tikz\draw[blue,thick,dashed] (0,0) circle (.3);}
\end{overpic}}&
{\begin{overpic}[width=.25\textwidth]{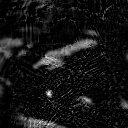}
\put(30,16) {\tikz\draw[red,thick,dashed] (0,0) circle (.3);}
\put(25,40){\tikz\draw[blue,thick,dashed] (0,0) circle (.3);}
\end{overpic}}& 
\includegraphics[width=.1\textwidth]{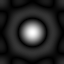}
\\
(a)&(b)&(c)&\\
{\begin{overpic}[width=.25\textwidth]{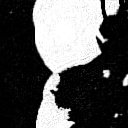}
\put(8,25) {\tikz\draw[red,thick,dashed] (0,0) circle (.3);}
\put(35,40){\tikz\draw[blue,thick,dashed] (0,0) circle (.3);}
\end{overpic}}&
{\begin{overpic}[width=.25\textwidth]{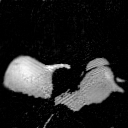}
\put(8,25) {\tikz\draw[red,thick,dashed] (0,0) circle (.3);}
\put(55,15){\tikz\draw[blue,thick,dashed] (0,0) circle (.3);}
\end{overpic}}&
{\begin{overpic}[width=.25\textwidth]{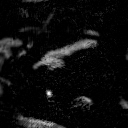}
\put(30,16) {\tikz\draw[red,thick,dashed] (0,0) circle (.3);}
\put(25,40){\tikz\draw[blue,thick,dashed] (0,0) circle (.3);}
\end{overpic}}& 
\includegraphics[width=.1\textwidth]{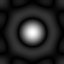}
\\
(d)&(e)&(f)&\\
{\begin{overpic}[width=.25\textwidth]{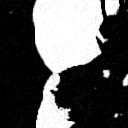}
\put(8,25) {\tikz\draw[red,thick,dashed] (0,0) circle (.3);}
\put(35,40){\tikz\draw[blue,thick,dashed] (0,0) circle (.3);}
\end{overpic}}&
{\begin{overpic}[width=.25\textwidth]{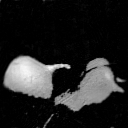}
\put(8,25) {\tikz\draw[red,thick,dashed] (0,0) circle (.3);}
\put(55,15){\tikz\draw[blue,thick,dashed] (0,0) circle (.3);}
\end{overpic}}&
{\begin{overpic}[width=.25\textwidth]{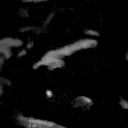}\qquad
\put(30,16) {\tikz\draw[red,thick,dashed] (0,0) circle (.3);}
\put(25,40){\tikz\draw[blue,thick,dashed] (0,0) circle (.3);}
\end{overpic}}& 
\includegraphics[width=.1\textwidth]{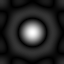}
\\
(g)&(h)&(i)&
\end{tabular}
\caption{Reconstruction results using a known dictionary of spectra from Poisson noisy data with SNR=29.2 dB and scan stepsize=32.
(a)-(c): Standard two-step method;
(d)-(f): SPA without regularization; (g)-(i) SPA with TV regularization. The recovered probes are put on the right column for the two-step method, SPA, and SPA with TV (from top to bottom) respectively.}
\end{center}
\label{fig2}
\end{figure}

\begin{figure}
\begin{center}
\begin{tabular}{cccc}
{\begin{overpic}[width=.25\textwidth]{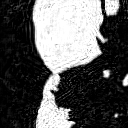}
\put(45,20) {\tikz\draw[red,thick,dashed] (0,0) circle (.3);}
\put(35,40){\tikz\draw[blue,thick,dashed] (0,0) circle (.3);}
\end{overpic}}&
{\begin{overpic}[width=.25\textwidth]{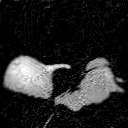}
\put(24,20) {\tikz\draw[red,thick,dashed] (0,0) circle (.3);}
\put(55,15){\tikz\draw[blue,thick,dashed] (0,0) circle (.3);}
\end{overpic}} &
{\begin{overpic}[width=.25\textwidth]{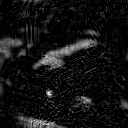}
\put(30,16) {\tikz\draw[red,thick,dashed] (0,0) circle (.3);}
\put(60,60){\tikz\draw[blue,thick,dashed] (0,0) circle (.3);}
\end{overpic}}& 
\includegraphics[width=.1\textwidth]{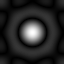}
\\
(a)&(b)&(c)\\
{\begin{overpic}[width=.25\textwidth]{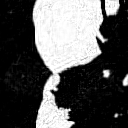}
\put(45,20) {\tikz\draw[red,thick,dashed] (0,0) circle (.3);}
\put(35,40){\tikz\draw[blue,thick,dashed] (0,0) circle (.3);}
\end{overpic}}&
{\begin{overpic}[width=.25\textwidth]{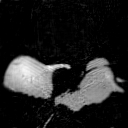}
\put(24,20) {\tikz\draw[red,thick,dashed] (0,0) circle (.3);}
\put(55,15){\tikz\draw[blue,thick,dashed] (0,0) circle (.3);}
\end{overpic}}&
{\begin{overpic}[width=.25\textwidth]{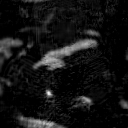}
\put(30,16) {\tikz\draw[red,thick,dashed] (0,0) circle (.3);}
\put(60,60){\tikz\draw[blue,thick,dashed] (0,0) circle (.3);}
\end{overpic}}& 
\includegraphics[width=.1\textwidth]{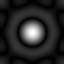}
\\
(d)&(e)&(f)&\\
{\begin{overpic}[width=.25\textwidth]{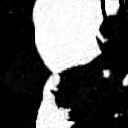}
\put(45,20) {\tikz\draw[red,thick,dashed] (0,0) circle (.3);}
\put(35,40){\tikz\draw[blue,thick,dashed] (0,0) circle (.3);}
\end{overpic}}&
{\begin{overpic}[width=.25\textwidth]{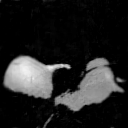}
\put(24,20) {\tikz\draw[red,thick,dashed] (0,0) circle (.3);}
\put(55,15){\tikz\draw[blue,thick,dashed] (0,0) circle (.3);}
\end{overpic}}&
{\begin{overpic}[width=.25\textwidth]{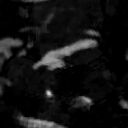}
\put(30,16) {\tikz\draw[red,thick,dashed] (0,0) circle (.3);}
\put(60,60){\tikz\draw[blue,thick,dashed] (0,0) circle (.3);}
\end{overpic}}& 
\includegraphics[width=.1\textwidth]{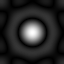}
\\
(g)&(h)&(i)&
\end{tabular}
\end{center}
\caption{
Reconstruction results using a partially known dictionary of spectra, from Poisson noisy data with SNR=29.2 dB and scan stepsize=32.
(a)-(c): Standard two-step method;
(d)-(f): SPA without regularization; (g)-(i) SPA with TV regularization. The recovered probes are put on the right column for the two-step method, SPA, and SPA with TV (from top to bottom) respectively
}
\label{fig3}
\end{figure}

The first simulation assesses the reconstruction quality achieved by the proposed SPA algorithm, compared with the two-step method, when using a scan stepsize of 32 pixels. Figs.~\ref{fig2}-\ref{fig3} depict the reconstructed images when using complete and incomplete dictionaries, respectively. The SPA simulations are performed without and with regularization, in rows 2 and 3 of Figs. \ref{fig2}-\ref{fig3}, respectively. Visually, we can see obvious artifacts in the recovered images when using the two-step method (first row of Figs. \ref{fig2}-\ref{fig3}). Such artifacts are greatly enhanced when reconstructed using SPA. Specifically, clear improvements can be identified in the regions corresponding to the red and blue circles for all three materials in both Fig.~\ref{fig2} and~\ref{fig3}.
The SNRs of the recovery results parallel the visual analysis. For the completely known dictionary, the two-step method achieves a SNR of 14.0 dB for the above simulation, whereas SPA achieves 18.1 dB (no regularization), and 18.8 dB (regularization). In the partially known dictionary simulation, the SNRs are 13.8, 15.8, and 16.7 dBs, for the two-step method and SPA with no-regularization and regularization, respectively, which achieves a comparative gain of more than 2 dB, similarly as in the known dictionary case.

The phase ambiguity is an inherent problem of the two-step method that causes a loss in reconstruction accuracy. For example, for the simulation shown in Fig.~\ref{fig2}, the SNR without phase correction is only 12.3 dB, reaching 14.0 dB after applying correction.  Even when using effective phase correction post-process, SPA proves to be more efficient for the simulations performed: higher quality reconstruction are achieved overall, and there is no need to correct the phase ambiguity due to the iterative reconstruction exploiting the low-rank structure and positivity constraint of the thickness function. 

%\graphicspath{images/}
%Stepsize=32. Frame size is $64\times 64$. 

\subsection{Robustness and convergence}
\onecolumn

\begin{figure}
\begin{center}
\begin{tabular}{cccc}
{\begin{overpic}[width=.25\textwidth]{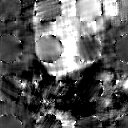}
\put(35,20) {\tikz\draw[red,thick,dashed] (0,0) circle (.3);}
\put(60,55){\tikz\draw[blue,thick,dashed] (0,0) circle (.3);}
\end{overpic}}&
{\begin{overpic}[width=.25\textwidth]{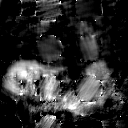}
\put(24,20) {\tikz\draw[red,thick,dashed] (0,0) circle (.3);}
\put(55,15){\tikz\draw[blue,thick,dashed] (0,0) circle (.3);}
\end{overpic}}&
{\begin{overpic}[width=.25\textwidth]{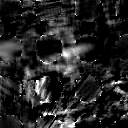}
\put(30,16) {\tikz\draw[red,thick,dashed] (0,0) circle (.3);}
\put(40,40){\tikz\draw[blue,thick,dashed] (0,0) circle (.3);}
\end{overpic}}& 
\includegraphics[width=.1\textwidth]{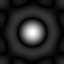}
\\
(a)&(b)&(c)&\\
{\begin{overpic}[width=.25\textwidth]{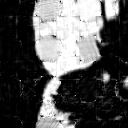}
\put(35,20) {\tikz\draw[red,thick,dashed] (0,0) circle (.3);}
\put(60,55){\tikz\draw[blue,thick,dashed] (0,0) circle (.3);}
\end{overpic}}&
{\begin{overpic}[width=.25\textwidth]{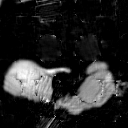}
\put(24,20) {\tikz\draw[red,thick,dashed] (0,0) circle (.3);}
\put(55,15){\tikz\draw[blue,thick,dashed] (0,0) circle (.3);}
\end{overpic}}&
{\begin{overpic}[width=.25\textwidth]{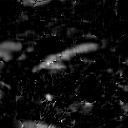}
\put(30,16) {\tikz\draw[red,thick,dashed] (0,0) circle (.3);}
\put(40,40){\tikz\draw[blue,thick,dashed] (0,0) circle (.3);}
\end{overpic}}& 
\includegraphics[width=.1\textwidth]{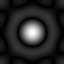}\\
(d)&(e)&(f)&\\
{\begin{overpic}[width=.25\textwidth]{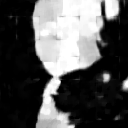}
\put(35,20) {\tikz\draw[red,thick,dashed] (0,0) circle (.3);}
\put(60,55){\tikz\draw[blue,thick,dashed] (0,0) circle (.3);}
\end{overpic}}&
{\begin{overpic}[width=.25\textwidth]{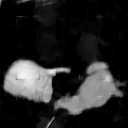}
\put(24,20) {\tikz\draw[red,thick,dashed] (0,0) circle (.3);}
\put(55,15){\tikz\draw[blue,thick,dashed] (0,0) circle (.3);}
\end{overpic}}&
{\begin{overpic}[width=.25\textwidth]{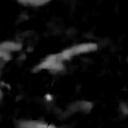}
\put(30,16) {\tikz\draw[red,thick,dashed] (0,0) circle (.3);}
\put(40,40){\tikz\draw[blue,thick,dashed] (0,0) circle (.3);}
\end{overpic}}& 
\includegraphics[width=.1\textwidth]{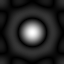}
\\
(g)&(h)&(i)&\\
%\subfloat[][]{\includegraphics[width=.25\textwidth]{eg1two_0}}
%\subfloat[][]{\includegraphics[width=.25\textwidth]{eg1two_1}}
%\subfloat[][]{\includegraphics[width=.25\textwidth]{eg1two_2}}\\
%\subfloat[][]{\includegraphics[width=.25\textwidth]{eg1HSP_0}}
%\subfloat[][]{\includegraphics[width=.25\textwidth]{eg1HSP_1}}
%\subfloat[][]{\includegraphics[width=.25\textwidth]{eg1HSP_2}}\\
%\subfloat[][]{\includegraphics[width=.25\textwidth]{eg1HSP_r0}}
%\subfloat[][]{\includegraphics[width=.25\textwidth]{eg1HSP_r1}}
%\subfloat[][]{\includegraphics[width=.25\textwidth]{eg1HSP_r2}}
\end{tabular}
\caption{
Reconstruction results using a known dictionary of spectra, from Poisson noisy data  with SNR=29.0 dB and scan stepsize=40.
(a)-(c): Standard two-step method;
(d)-(f): SPA without regularization; (g)-(i) SPA with TV regularization. The recovered probes are put on the right column for the two-step method, SPA, and SPA with TV (from top to bottom) respectively}
\end{center}
\label{fig4}
\end{figure}

\begin{table}
%\vskip -.02in
    \caption{SNR in dB from reconstruction results with different scan stepsizes when using the two-step method, SPA and SPA with TV regularization.}
%    \vskip .1in
    \begin{center}
    \begin{tabular}{|c|c|c|c|c|}
    \hline
    \multicolumn{2}{|c|}{stepsize} & 36&38&40 \\
          \hline
         \multirow{3}{*}{ SNR}&two-step    & 11.5&7.0   &3.4 \\
                            &SPA           & 15.1&15.5  & 12.3  \\
                                  &SPA + TV&16.0&16.0  & 13.8\\
      \hline
    \end{tabular}
    \end{center}
    \label{tab0}
%    \vskip -.1in
\end{table}

The following simulation assesses the robustness of the proposed algorithm when varying the scanning stepsizes.  The quantitative results of this simulation are presented in Table \ref{tab0}. The results demonstrate the enhanced robustness of SPA when handling bigger stepsizes, compared to the reference two-step method, achieving up to 10 dB increase in SNR. To permit a better visual analysis, we provide the reconstruction results of the three algorithms with 40 pixels stepsize in Fig.~\ref{fig4}. The figure highlights the dramatic improvement achieved by SPA compared with the standard two-step method when reconstructing low-redundancy ptychographic data. Specifically, we can see how the features within the blue and red circles are almost lost in the two-step reconstruction, while they can be clearly observed when reconstructing using SPA.

%\subsection{Further discussion on the Spectrum  dictionaries}
{ Generally speaking, to make the proposed algorithms work, the basic condition is to assume $D$ has full row rank such that $DD^*$ is non-singular. However, the performance should also rely on the similarity of spectral elements. Here  we introduce a factor $s\in[0, 0.5]$ to generate a new dictionary $D_s\in\mathbb C^{3\times10}$ with $D_s[1,k]=(1-s)D[1,k]+sD[2,k], D_s[2,k]=(1-s)D[2,k]+sD[1,k] \forall 1\leq k\leq 10.$ Readily one knows that (1) $D_0=D$; and (2) the first two rows are exactly the same if $s=1/2$ ($D_{1/2}$ does not have full row rank). Please see Fig. \ref{fig8} for the dictionaries with $s=0.1, 0.3,$ and $0.45$.  Therefore, the parameter $s$ can be used to control the similarity of the new spectral  dictionary (bigger $s$ implies higher similarity). We test the impact of proposed algorithm  SPA by the different similarity of spectral dictionaries, and please see reconstruction results in Fig. \ref{fig6-1} with $s=0.1, 0.3, 0.45$, and SNR changes in Fig. \ref{fig6} with $s\in\{ 0, 0.1,  0.2,0.3, 0.4, 0.45,0.475,0.49\}$. One readily knows that the quality of  reconstruction results by proposed SPA decays as the spectral becomes similar (the parameter $s$ gets close to $0.5$). Hence, to get better reconstruction, one should design the experiments with little similarity in spectral dictionaries.

\begin{figure}
\begin{center}
\begin{tabular}{cc}%\centering
\includegraphics[width=.45\textwidth]{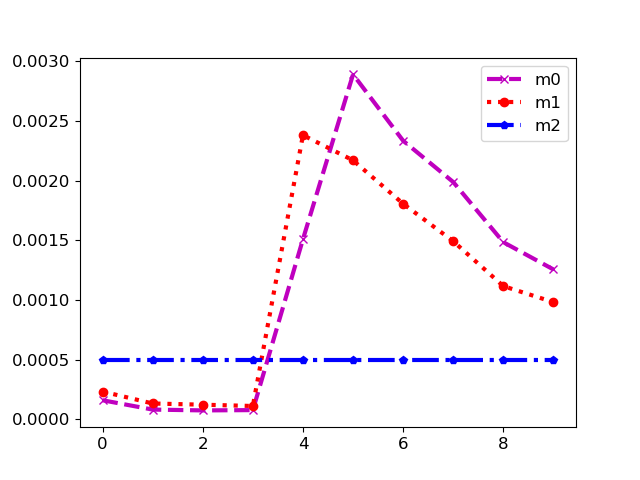} &
\includegraphics[width=.45\textwidth]{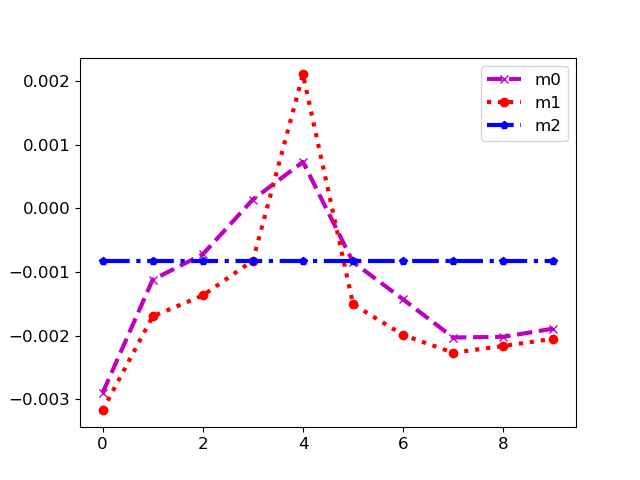}\\
\multicolumn{2}{c}{$s=0.1$}\\
\includegraphics[width=.45\textwidth]{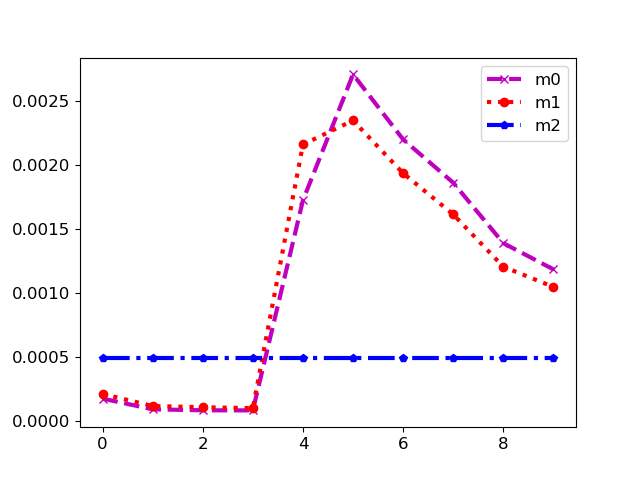} &
\includegraphics[width=.45\textwidth]{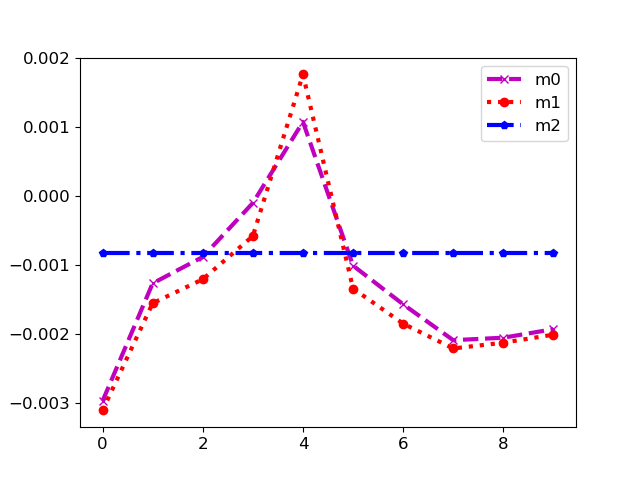}\\
\multicolumn{2}{c}{$s=0.3$}\\
\includegraphics[width=.45\textwidth]{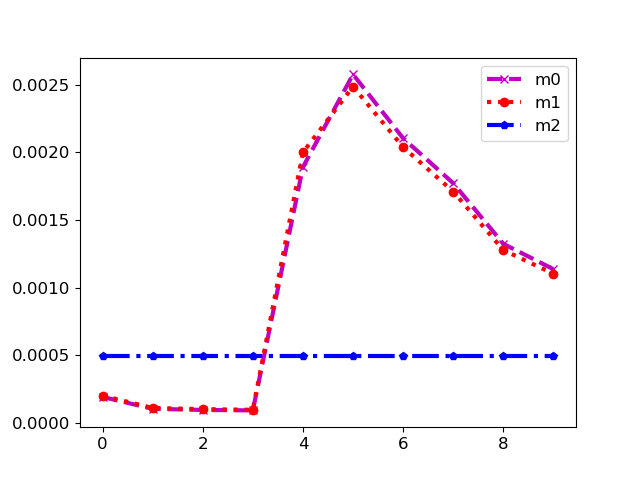} &
\includegraphics[width=.45\textwidth]{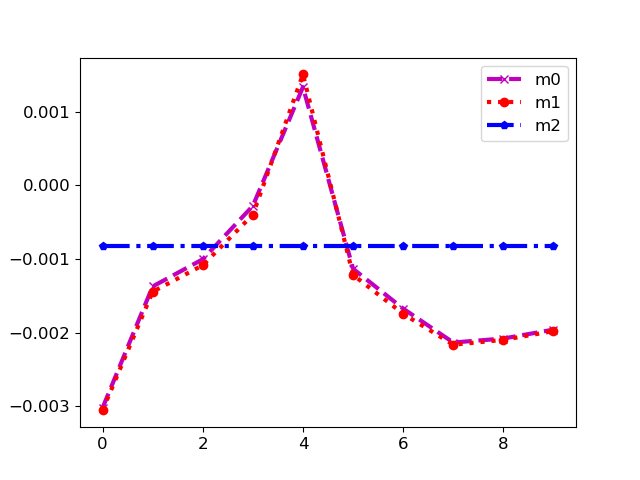}\\
\multicolumn{2}{c}{$s=0.45$}
\end{tabular}
\end{center}
\caption{Synthetic spectrum dictionaries $D_s$ for different $s$ (three different materials m0 (PMMA), m1 (PS) and m2 (constant)), with real part (left) and imaginary part (right). x- and y-axis denote spectrum and different energies, respectively. }
\label{fig8}
\end{figure}

\begin{figure}
    \centering
    \includegraphics[width=.5\textwidth]{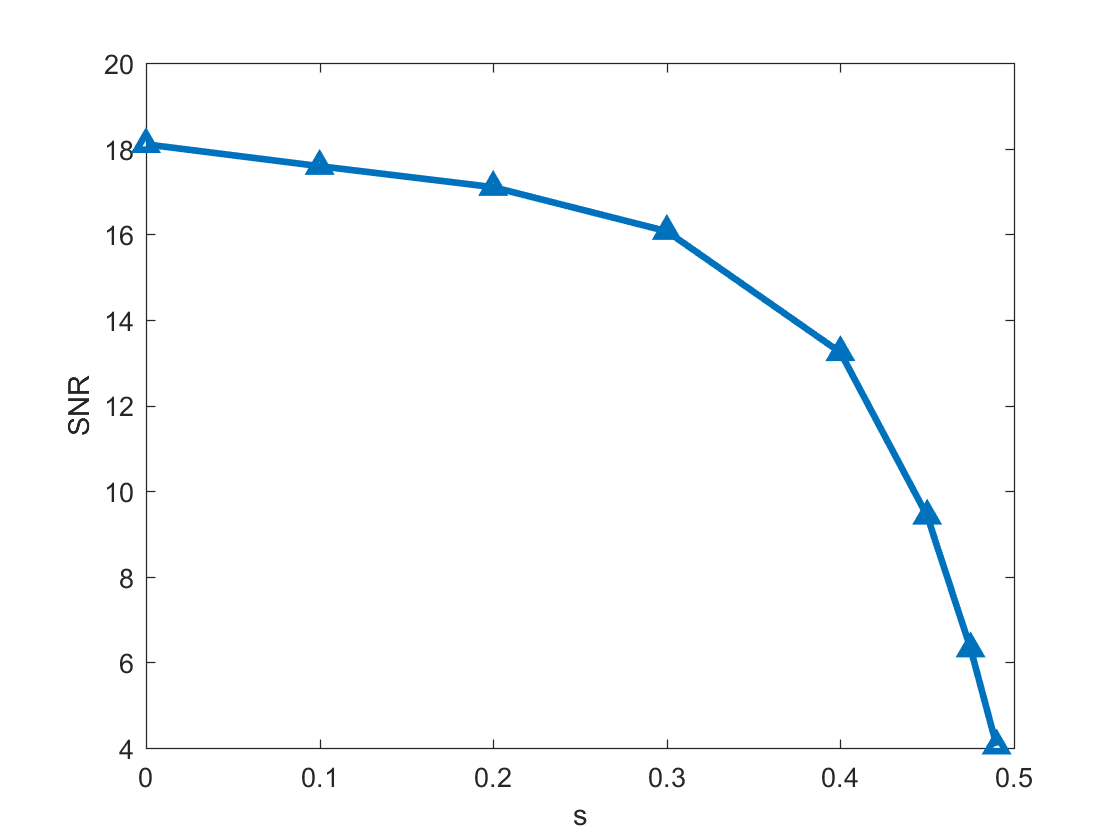}
    \caption{SNR changes v.s. similarity parameter $s$.}
    \label{fig6}
\end{figure}

\begin{figure}
\begin{center}
\begin{tabular}{ccc}
\includegraphics[width=.24\textwidth]{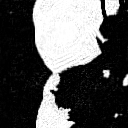}&
\includegraphics[width=.24\textwidth]{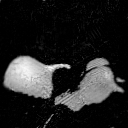}&
\includegraphics[width=.24\textwidth]{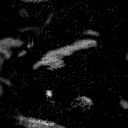} \\
\multicolumn{3}{c}{$s=0.1$}\\
\includegraphics[width=.24\textwidth]{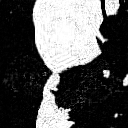}&
\includegraphics[width=.24\textwidth]{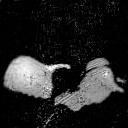}&
\includegraphics[width=.24\textwidth]{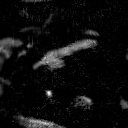} \\
\multicolumn{3}{c}{$s=0.3$}\\
\includegraphics[width=.24\textwidth]{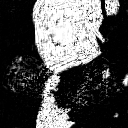}&
\includegraphics[width=.24\textwidth]{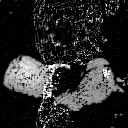}&
\includegraphics[width=.24\textwidth]{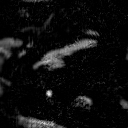} \\
\multicolumn{3}{c}{$s=0.45$}\\
\end{tabular}
\caption{
Reconstruction results by  proposed SPA using different dictionaries of spectra ($s=0.1, 0.3, 0.45$), from Poisson noisy data  with SNR=29.2 dB and scan stepsize=32.}
\end{center}
\label{fig6-1}
\end{figure}
}

\begin{figure}
\begin{center}
\begin{tabular}{cc}
{\includegraphics[width=.35\textwidth]{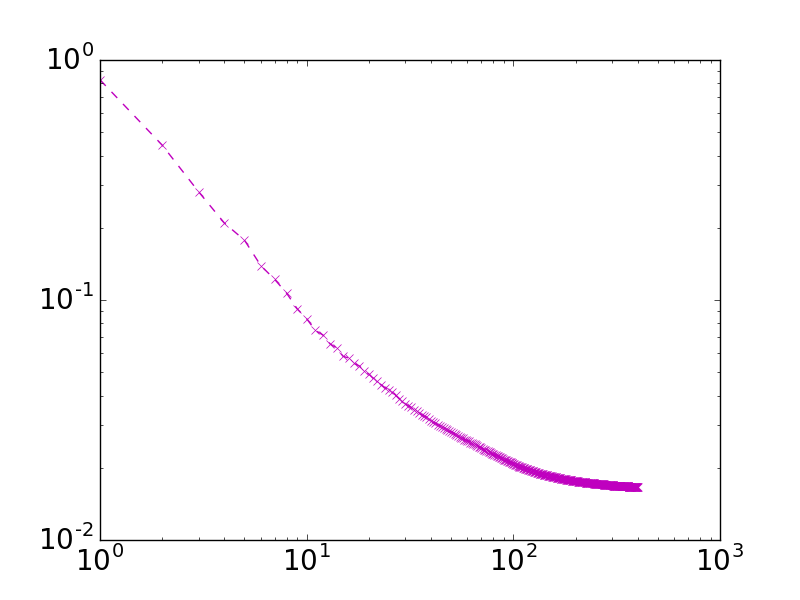}}&
{\includegraphics[width=.35\textwidth]{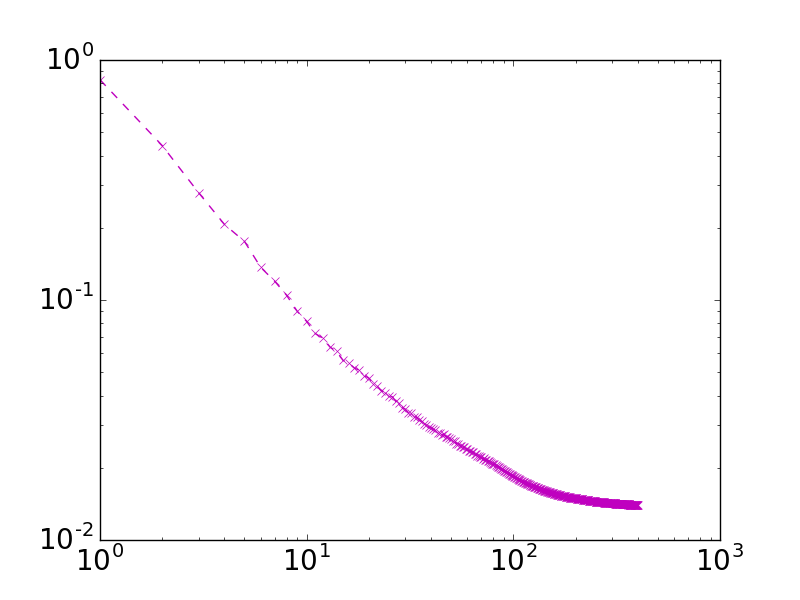}}\\
(a)&(b)
\end{tabular}\\
\caption{Error $\tfrac{\|x^k-X^{k-1}\|}{\|X^k\|}$ variation v.s. iteration number for SPA without (a) and with TV regularization (b). X- and y-axis denote iteration numbers and errors, respectively.}
\label{fig5}
\end{center}
\end{figure}

The last simulation depicts the error curve achieved by the SPA algorithm, shown in Fig.~\ref{fig5}. The results demonstrate an steady decrease of the successive errors of the proposed algorithm, both with and without regularization. 

%%%%%%%%%% If using BibTeX:
\section{Conclusions}
\onecolumn
This paper presents the first iterative spectroscopy ptychography solution. The proposed SPA algorithm is based on a novel spectro-ptychography model and it is constructed considering both a completely known and partially known dictionaries. Numerical simulations show that SPA produces more accurate results with clearer features compared with the standard two-step method. In the future, we will extend our work to thicker samples, where the first-order Taylor expansion is not sufficiently accurate. We also plan to investigate the use of Kramers-Kronig relationships \cite{hirose2017use}, explore the case using a completely unknown dictionary,
and further provide software for real experimental data analysis.

\section{Acknowledgments}
\onecolumn
This work of the first author was partially supported by National Natural Science Foundation of China (Nos.11871372,
11501413), Natural Science Foundation of Tianjin (No.18JCYBJC16600), 2017-Outstanding Young Innovation Team Cultivation Program (No.043-135202TD1703) and Innovation Project (No.043-135202XC1605) of Tianjin Normal University, Tianjin
Young Backbone of Innovative Personnel Training Program and Program for Innovative Research Team in Universities of
Tianjin (No.TD13-5078). This work was also partially funded by the Advanced Light Source and the Center for Advanced Mathematics for Energy Research Applications, a joint ASCR-BES funded project within the Office of Science, US Department of Energy, under contract number DOE-DE-AC03-76SF00098.

\vskip 2in
\appendix 
\noindent {\bf{Appendix}}\\

%\begin{minipage}{.95\linewidth}
{
\vskip.1in
\hskip .2in
%\hrule \vskip .05in
\hrule \vskip .1in
\centering  \textbf{Algorithm 1: SPA} \vskip .1in
\hrule\vskip .05in}
\begin{itemize}
\item[0.] Initialization:   Compute the SVD of $\hat D\hat D^*$  as $\hat D\hat D^*=V\mathscr S V^*.$ Set $Y^0:=\mathbf 1,  \omega^0:=\mathcal F^*\left(\tfrac{\sumMy\nolimits_{l,j} \sqrt{I_{l,j}}}{L\times J}\right), Z^0_{l}:=\mathcal A(\omega^0,  Y^0_{l}), \Lambda=\bm 0,$ and $\Gamma=\bm 0.$
Set $k:=0.$
\item[1.] Update the probe $\omega^{k+1}$ by one-step projected gradient descent method
\begin{equation}
\begin{split}
\omega^{k+1}=\mathrm{Proj}_{\mathscr W}\left(\tfrac{\gamma_1\omega^k}{(\sumMy\nolimits_{l,j}|\mathcal S_j Y^k_l|^2) +\gamma_1 \mathbf 1}+\tfrac{\sumMy\nolimits_{l,j} \mathcal F^*(Z^k_{l,j}+\Lambda^k_{l,j})\circ  \mathcal S_j  (Y^k_l)^*}{(\sumMy\nolimits_{l,j}|\mathcal S_j Y^k_l|^2) +\gamma_1 \mathbf 1}\right),
\end{split}
\end{equation}
with $\gamma_1=0.1 \times\|\sum\nolimits_{l,j}|\mathcal S_j Y^k_l|^2\|_\infty.$
\item[2.] Update the thickness function $X^{k+1}$ by
\[
X_c^{k+1}=\max\{0, \mathrm{Denoise}_{\delta/\beta}(\Re(((Y^k-\bm 1)\hat D-\Gamma^k))_c)\}~\forall 0\leq c\leq C-1.\]
\item[3.] Update $Y^{k+1}$
by
\begin{equation}
\label{eq:solverY}
Y^{k+1}=\hat Y^{k+1} V^*,~~\hat Y^{k+1}_l=\dfrac{((\lambda Q^k+\gamma^k_2 Y^k+\beta(\Gamma^k+X^{k+1}+\bm 1\hat D)\hat D^*)V)_l}{(\lambda\sumMy\nolimits_j |\mathcal S^T_j \omega^{k+1}|^2+\gamma^k_2 \mathbf 1)+\beta\mathscr S_{l,l}\bm 1}~~\forall~0\leq l\leq L-1.
\end{equation}
%by \eqref{slv1}-\eqref{slv2}, which solves the following Sylvester equation:
%\[
%\begin{split}
%\mathrm{diag}(\lambda\sum\nolimits_j |\mathcal S^T_j \omega^{k+1}|^2+\gamma^k_2 \mathbf 1)Y+\beta  Y \hat %D\hat D^*=\lambda Q^k+\gamma_2 Y^k+\beta(\Gamma+X^{k+1}+\bm 1\hat D)\hat D^*,
%\end{split}
%\]
with $ Q^k:=(Q^k_0, Q^k_1,\ldots,Q^k_{L-1})\in\mathbb C^{N, L},$ $Q_l^k:=\sum\nolimits_j\mathcal S_j^T ((\omega^{k+1})^*\circ \mathcal F^*(\Lambda_l^k+Z^k_l)),$ and $\gamma_2^k=0.1\times\lambda \|\sum\nolimits_j |\mathcal S_j \omega^{k+1}|^2\|_\infty.$
\item[4.] Update the auxiliary variable $Z^{k+1}$ by
\[
Z_l^{k+1}=\tfrac{\sqrt{4(1+\lambda)I_l+\lambda^2|\hat Z^k_l|^2}+\lambda|\hat Z^k_l|}{2(1+\lambda)}\circ\mathrm{sign}(\hat Z^k_l),\]
with $\hat Z^k_l:=\mathcal A(\omega^{k+1},Y^{k+1}_l)-\Lambda^k_l$.
\item[5.] Update the multipliers $\Lambda$ and $\Gamma$ by
\[
\begin{split}
&\Lambda_l\leftarrow \Lambda_l+Z_l-\mathcal A(\omega, \mathcal T_l Y_l)~\forall 0\leq l\leq L-1;\\
&\Gamma   \leftarrow  \Gamma+ X-(Y-\mathbf 1)\hat D.
\end{split}
\]
\item[6.] When satisfying the stopping condition, output $X^{k+1}$ as the final thickness, otherwise, go to Step 1.
\end{itemize}
\hrule \vskip .2in
%\end{minipage}

\vskip .2in
%\begin{minipage}{.95\linewidth}
{
\vskip.1in
\hskip .2in
%\hrule \vskip .05in
\hrule \vskip .1in
\centering  \textbf{Algorithm 2: SPA with incomplete dictionary} \vskip .1in
\hrule\vskip .05in}
\begin{itemize}
\item[0.] Initialization:   Compute the SVD of $\hat D_r\hat D_r^*$  as $\hat D_r\hat D_r^*=V_r\mathscr S_r V_r^*.$ Set $Y^0:=\mathbf 1,  \omega^0:=\mathcal F^*\left(\tfrac{\sumMy\nolimits_{l,j} \sqrt{I_{l,j}}}{L\times J}\right), Z^0_{l}:=\mathcal A(\omega^0,  Y^0_{l}), \Lambda=\bm 0,$ and $\Gamma_r=\bm 0.$
Set $k:=0.$
\item[1.] Update the probe $\omega^{k+1}$ as Step 1 of Algorithm 1.
\item[2.] Update the thickness function $X^{k+1}$ by
\[
X_c^{k+1}=\max\{0, \mathrm{Denoise}_{\delta/\beta}((\Re(Y^k-\bm 1)\hat D_r-\Gamma^k_r)_c)\}~\forall 0\leq c\leq C-1.\]
\item[3.] Update $Y^{k+1}=Y_r^{k+1}+\mathrm i\times Y_i^{k+1}$ (Imaginary unit $\mathrm i:=\sqrt{-1}$) with 
real part update by
\begin{equation}
\label{eq:solverYIm}
\begin{split}
Y^{k+1}_r&=\hat Y_r^{k+1} V_r^*,\\
(\hat Y_r^{k+1})_l&=\dfrac{((\lambda \Re(Q^k)+\hat \gamma^k_2 \Re(Y^k)+\beta(\Gamma_r^k+X^{k+1}+\bm 1\hat D_r)\hat D_r^*)V)_l}{(\lambda\sumMy\nolimits_j |\mathcal S^T_j \omega^{k+1}|^2+\gamma^k_2 \mathbf 1)+\beta\mathscr S_{l,l}\bm 1}, \\
&\forall~0\leq l\leq L-1,
\end{split}
\end{equation}
with  $\hat\gamma_2^k=0.1\times\lambda \|\sum\nolimits_j |\mathcal S_j \omega^{k+1}|^2\|_\infty,$
and imaginary part update by 
\[
Y^{k+1}_i=\frac{\lambda \Im(Q^k)+\hat\gamma_2 \Im(Y^{k})}{\lambda\sum\nolimits_j |\mathcal S^T_j \omega^{k+1}|^2+\hat\gamma_2 \mathbf 1}.
\]

\item[4.] Update the auxiliary variable $Z^{k+1}$ as Step 4 of Algorithm 1.
\item[5.] Update the multipliers $\Lambda$ and $\Gamma_r$ by
\[
\begin{split}
&\Lambda_l\leftarrow \Lambda_l+Z_l-\mathcal A(\omega, \mathcal T_l Y_l)~\forall 0\leq l\leq L-1;\\
&\Gamma_r   \leftarrow  \Gamma_r+ X-\Re(Y-\mathbf 1)\hat D_r.
\end{split}
\]
\item[6.] When satisfying the stopping condition, output $X^{k+1}$ as the final thickness, otherwise, go to Step 1.
\end{itemize}
\hrule \vskip .2in
%\end{minipage}

\
\ifx \aff \undefined
\else
 \bibliographystyle{iucr}    
\fi

%\referencelist{rD}

%\bibliography{rD}

%\bibliographystyle{siam}
\bibliography{rD}

@book{stohr2013nexafs,
  title={{NEXAFS} spectroscopy},
  author={St{\"o}hr, Joachim},
  volume={25},
  year={2013},
  publisher={Springer Science \& Business Media}
}

@book{koningsberger1988x,
title = "X-ray absorption : principles, applications, techniques of {EXAFS}, {SEXAFS} and {XANES}",
author = "D.C. Koningsberger and R. Prins",
year = "1988",
language = "English",
isbn = "0-471-87547-3",
series = "Chemical analysis",
publisher = "Wiley-Interscience",
}

@inproceedings{chang2019iterative,
  title={Iterative Joint Ptychography-Tomography with Total Variation Regularization},
  author={Chang, Huibin and Enfedaque, Pablo and Marchesini, Stefano},
  booktitle={2019 IEEE International Conference on Image Processing (ICIP)},
  pages={2931--2935},
  year={2019},
  organization={IEEE}
}

@article{henke1993x,
  title={X-ray interactions: photoabsorption, scattering, transmission, and reflection at E= 50-30,000 eV, Z= 1-92},
  author={Henke, Burton L and Gullikson, Eric M and Davis, John C},
  journal={Atomic data and nuclear data tables},
  volume={54},
  number={2},
  pages={181--342},
  year={1993},
  publisher={Elsevier}
}

@article{yan2013accurate,
  title={Accurate and Facile Determination of the Index of Refraction of Organic Thin Films Near the Carbon 1 s Absorption Edge},
  author={Yan, Hongping and Wang, Cheng and McCarn, Allison R and Ade, Harald},
  journal={Physical review letters},
  volume={110},
  number={17},
  pages={177401},
  year={2013},
  publisher={APS}
}

@article{kronig1926theory,
  title={On the theory of dispersion of x-rays},
  author={Kronig, R de L},
  journal={Josa},
  volume={12},
  number={6},
  pages={547--557},
  year={1926},
  publisher={Optical Society of America}
}

@article{hirose2017use,
  title={Use of Kramers--Kronig relation in phase retrieval calculation in X-ray spectro-ptychography},
  author={Hirose, Makoto and Shimomura, Kei and Burdet, Nicolas and Takahashi, Yukio},
  journal={Optics express},
  volume={25},
  number={8},
  pages={8593--8603},
  year={2017},
  publisher={Optical Society of America}
}

@article{aslan2019joint,
  title={Joint ptycho-tomography reconstruction through alternating direction method of multipliers},
  author={Aslan, Selin and Nikitin, Viktor and Ching, Daniel J and Bicer, Tekin and Leyffer, Sven and G{\"u}rsoy, Do{\u{g}}a},
  journal={Optics Express},
  volume={27},
  number={6},
  pages={9128--9143},
  year={2019},
  publisher={Optical Society of America}
}

@article{adams1986spectral,
  title={Spectral mixture modeling: A new analysis of rock and soil types at the Viking Lander 1 site},
  author={Adams, John B and Smith, Milton O and Johnson, Paul E},
  journal={Journal of Geophysical Research: Solid Earth},
  volume={91},
  number={B8},
  pages={8098--8112},
  year={1986},
  publisher={Wiley Online Library}
}

@article{yu2018three,
  title={Three-dimensional localization of nanoscale battery reactions using soft X-ray tomography},
  author={Yu, Young-Sang and Farmand, Maryam and Kim, Chunjoong and Liu, Yijin and Grey, Clare P and Strobridge, Fiona C and Tyliszczak, Tolek and Celestre, Rich and Denes, Peter and Joseph, John and others},
  journal={Nature communications},
  volume={9},
  number={1},
  pages={921},
  year={2018},
  publisher={Nature Publishing Group}
}

@article{farmand2017near,
  title={Near-edge X-ray refraction fine structure microscopy},
  author={Farmand, Maryam and Celestre, Richard and Denes, Peter and Kilcoyne, AL David and Marchesini, Stefano and Padmore, Howard and Tyliszczak, Tolek and Warwick, Tony and Shi, Xiaowen and Lee, James and others},
  journal={Applied Physics Letters},
  volume={110},
  number={6},
  pages={063101},
  year={2017},
  publisher={AIP Publishing}
}

@article{lerotic2004cluster,
  title={Cluster analysis of soft X-ray spectromicroscopy data},
  author={Lerotic, M and Jacobsen, C and Sch{\"a}fer, T and Vogt, S},
  journal={Ultramicroscopy},
  volume={100},
  number={1-2},
  pages={35--57},
  year={2004},
  publisher={Elsevier}
}

@article{rodenburg2007hard,
  title={Hard-x-ray lensless imaging of extended objects},
  author={Rodenburg, JM and Hurst, AC and Cullis, AG and Dobson, BR and Pfeiffer, F and Bunk, O and David, C and Jefimovs, K and Johnson, I},
  journal={Physical review letters},
  volume={98},
  number={3},
  pages={034801},
  year={2007},
  publisher={APS}
}

@article{marchesini2013augmented,
  title={Augmented projections for ptychographic imaging},
  author={Marchesini, Stefano and Schirotzek, Andre and Yang, Chao and Wu, Hau-{T}ieng and Maia, Filipe},
  journal={Inverse Problems},
  volume={29},
  number={11},
  pages={115009},
  year={2013},
  publisher={IOP Publishing}
}

@article{chambolle2011first,
  title={A first-order primal-dual algorithm for convex problems with applications to imaging},
  author={Chambolle, Antonin and Pock, Thomas},
  journal={Journal of mathematical imaging and vision},
  volume={40},
  number={1},
  pages={120--145},
  year={2011},
  publisher={Springer}
}

@article{Chan2013,
  author        = "R. H. Chan and M. Tao and X.M. Yuan",
  title         = "Constrained total variational deblurring models and fast algorithms based on alternating direction method of multipliers",
  journal       = "SIAM J. Imaging Sci.",
  volume        = "6",
  number="1",
  pages="680-697",
  year={2013}
}

@article{chang2016Total,
  title={Total Variation--Based Phase Retrieval for Poisson Noise Removal},
  author={Chang, Huibin and Lou, Yifei and Duan, Yuping and Marchesini, Stefano},
  journal={SIAM Journal on Imaging Sciences},
  volume={11},
  number={1},
  pages={24--55},
  year={2018},
  publisher={SIAM}
}

@article{rudin1992nonlinear,
  title={Nonlinear total variation based noise removal algorithms},
  author={Rudin, Leonid I and Osher, Stanley and Fatemi, Emad},
  journal={Physica D: Nonlinear Phenomena},
  volume={60},
  number={1},
  pages={259--268},
  year={1992},
  publisher={Elsevier}
}

@article{wu2011augmented,
  title={Augmented Lagrangian method for total variation restoration with non-quadratic fidelity},
  author={Wu, Chunlin and Zhang, Juyong and Tai, Xue-Cheng},
  journal={Inverse problems and imaging},
  volume={5},
  number={1},
  pages={237--261},
  year={2011},
  publisher={Citeseer}
}

@book{glowinski1989augmented,
  title={Augmented Lagrangian and operator-splitting methods in nonlinear mechanics},
  author={Glowinski, Roland and Le Tallec, Patrick},
  year={1989},
  publisher={Philadelphia, PA: SIAM}
}

@article{wen2012,
  author        = "Z. Wen and C. Yang and X. Liu and S. Marchesini",
  title         = "Alternating direction methods for classical and ptychographic phase retrieval",
  journal       = "Inverse Probl.",
  volume        = "28",
  number="11",
  pages="115010",
  year="2012"
}

@article{
Sylvester,
  title={Sur l’{\'e}quation en matrices px= xq},
  author={Sylvester, James Joseph},
  journal={CR Acad. Sci. Paris},
  volume={99},
  number={2},
  pages={67--71},
  year={1884}
}

@article{chapman1996phase,
  title={Phase-retrieval X-ray microscopy by Wigner-distribution deconvolution},
  author={Chapman, Henry N},
  journal={Ultramicroscopy},
  volume={66},
  number={3},
  pages={153--172},
  year={1996},
  publisher={Elsevier}
}

@article{horstmeyer2015solving,
  title={Solving ptychography with a convex relaxation},
  author={Horstmeyer, Roarke and Chen, Richard Y and Ou, Xiaoze and Ames, Brendan and Tropp, Joel A and Yang, Changhuei},
  journal={New journal of physics},
  volume={17},
  number={5},
  pages={053044},
  year={2015},
  publisher={IOP Publishing}
}

@article{thibault2012maximum,
  title={Maximum-likelihood refinement for coherent diffractive imaging},
  author={Thibault, P and Guizar-Sicairos, M},
  journal={New Journal of Physics},
  volume={14},
  number={6},
  pages={063004},
  year={2012},
  publisher={IOP Publishing}
}

@article{hesse2015proximal,
  title={Proximal Heterogeneous Block Implicit-Explicit Method and Application to Blind Ptychographic Diffraction Imaging},
  author={Hesse, Robert and Luke, D Russell and Sabach, Shoham and Tam, Matthew K},
  journal={SIAM Journal on Imaging Sciences},
  volume={8},
  number={1},
  pages={426--457},
  year={2015},
  publisher={SIAM}
}

@article{rodenburg2004phase,
  title={A phase retrieval algorithm for shifting illumination},
  author={Rodenburg, John M and Faulkner, Helen ML},
  journal={Applied physics letters},
  volume={85},
  number={20},
  pages={4795--4797},
  year={2004},
  publisher={AIP Publishing}
}

@article{thibault2009probe,
  title={Probe retrieval in ptychographic coherent diffractive imaging},
  author={Thibault, Pierre and Dierolf, Martin and Bunk, Oliver and Menzel, Andreas and Pfeiffer, Franz},
  journal={Ultramicroscopy},
  volume={109},
  number={4},
  pages={338--343},
  year={2009},
  publisher={Elsevier}
}

@article{nellist1995resolution,
  title={Resolution beyond the'information limit'in transmission electron microscopy},
  author={Nellist, PD and McCallum, BC and Rodenburg, JM},
  journal={Nature},
  volume={374},
  number={6523},
  pages={630},
  year={1995},
  publisher={Nature Publishing Group}
}

@article{maiden2009improved,
  title={An improved ptychographical phase retrieval algorithm for diffractive imaging},
  author={Maiden, Andrew M and Rodenburg, John M},
  journal={Ultramicroscopy},
  volume={109},
  number={10},
  pages={1256--1262},
  year={2009},
  publisher={Elsevier}
}

@article{shi2016soft,
  title={Soft x-ray ptychography studies of nanoscale magnetic and structural correlations in thin SmCo5 films},
  author={Shi, X and Fischer, P and Neu, V and Elefant, D and Lee, JCT and Shapiro, DA and Farmand, M and Tyliszczak, T and Shiu, H-W and Marchesini, S and others},
  journal={Applied Physics Letters},
  volume={108},
  number={9},
  pages={094103},
  year={2016},
  publisher={AIP Publishing}
}

@article{shapiro2014chemical,
  title={Chemical composition mapping with nanometre resolution by soft X-ray microscopy},
  author={Shapiro, David A and Yu, Young-Sang and Tyliszczak, Tolek and Cabana, Jordi and Celestre, Rich and Chao, Weilun and Kaznatcheev, Konstantin and Kilcoyne, AL David and Maia, Filipe and Marchesini, Stefano and others},
  journal={Nature Photonics},
  volume={8},
  number={10},
  pages={765--769},
  year={2014},
  publisher={Nature Research}
}

@article{beckers2011chemical,
  title={Chemical contrast in soft x-ray ptychography},
  author={Beckers, Mike and Senkbeil, Tobias and Gorniak, Thomas and Reese, Michael and Giewekemeyer, Klaus and Gleber, Sophie-Charlotte and Salditt, Tim and Rosenhahn, Axel},
  journal={Physical Review Letters},
  volume={107},
  number={20},
  pages={208101},
  year={2011},
  publisher={APS}
}

@article{chang2018partially,
  title={Partially coherent ptychography by gradient decomposition of the probe},
  author={Chang, Huibin and Enfedaque, Pablo and Lou, Yifei and Marchesini, Stefano},
  journal={Acta Crystallographica Section A: Foundations and Advances},
  volume={74},
  number={3},
  pages={157--169},
  year={2018},
  publisher={Wiley Online Library}
}

@article{hoppe2013high,
  title={High-resolution chemical imaging of gold nanoparticles using hard x-ray ptychography},
  author={Hoppe, R and Reinhardt, J and Hofmann, G and Patommel, J and Grunwaldt, J-D and Damsgaard, Christian Danvad and Wellenreuther, G and Falkenberg, G and Schroer, CG},
  journal={Applied Physics Letters},
  volume={102},
  number={20},
  pages={203104},
  year={2013},
  publisher={AIP}
}

@article{maiden2013soft,
  title={Soft X-ray spectromicroscopy using ptychography with randomly phased illumination},
  author={Maiden, AM and Morrison, GR and Kaulich, B and Gianoncelli, A and Rodenburg, JM},
  journal={Nature communications},
  volume={4},
  pages={1669},
  year={2013},
  publisher={Nature Publishing Group}
}

@article{odstrvcil2018iterative,
  title={Iterative least-squares solver for generalized maximum-likelihood ptychography},
  author={Odstr{\v{c}}il, Michal and Menzel, Andreas and Guizar-Sicairos, Manuel},
  journal={Opt. Express},
  volume={26},
  number={3},
  pages={3108--3123},
  year={2018},
  publisher={Optical Society of America}
}

@article{chang2018Blind,
  author = {Chang, H. and Enfedaque, P. and Marchesini, S.},
title = {Blind Ptychographic Phase Retrieval via Convergent Alternating Direction Method of Multipliers},
journal = {SIAM J. Imaging Sci.},
volume = {12},
number = {1},
pages = {153-185},
year = {2019},
doi = {10.1137/18M1188446}
}

@article{vaswani2017low,
  title={Low-rank phase retrieval},
  author={Vaswani, Namrata and Nayer, Seyedehsara and Eldar, Yonina C},
  journal={IEEE Transactions on Signal Processing},
  volume={65},
  number={15},
  pages={4059--4074},
  year={2017},
  publisher={IEEE}
}

@inproceedings{chen2018low,
  title={Low rank fourier ptychography},
  author={Chen, Zhengyu and Jagatap, Gauri and Nayer, Seyedehsara and Hegde, Chinmay and Vaswani, Namrata},
  booktitle={2018 IEEE International Conference on Acoustics, Speech and Signal Processing (ICASSP)},
  pages={6538--6542},
  year={2018},
  organization={IEEE}
}

@article{liu2019low,
  title={Low-Rank Phase Retrieval via Variational Bayesian Learning},
  author={Liu, Kaihui and Wang, Jiayi and Xing, Zhengli and Yang, Linxiao and Fang, Jun},
  journal={IEEE Access},
  volume={7},
  pages={5642--5648},
  year={2019},
  publisher={IEEE}
}

@article{Simoncini2016,
title={Computational Methods for Linear Matrix Equations},
author={V. Simoncini},
journal={SIAM Review},
year={2016}, 
vol= {58},
pages={377-441}
}
%%%%%%%%%% If preparing manually:
% \newcommand{\enquote}[1]{``#1''}

% \bibitem{Zhang:14}
% Y.~Zhang, S.~Qiao, L.~Sun, Q.~W. Shi, W.~Huang, L.~Li, and Z.~Yang,
%   \enquote{Photoinduced active terahertz metamaterials with nanostructured
%   vanadium dioxide film deposited by sol-gel method,}
%   {\protect\JournalTitle{Optics Express}} \textbf{22}, 11070--11078 (2014).

% {Optical Society}, \enquote{{OSA Publishing},}
%   \url{http://www.osapublishing.org}.

% \bibitem{FORSTER2007}
% P.~Forster, V.~Ramaswamy, P.~Artaxo, T.~Bernsten, R.~Betts, D.~Fahey,
%   J.~Haywood, J.~Lean, D.~Lowe, G.~Myhre, J.~Nganga, R.~Prinn, G.~Raga,
%   M.~Schulz, and R.~V. Dorland, \enquote{Changes in atmospheric consituents and
%   in radiative forcing,} in \enquote{Climate Change 2007: The Physical Science
%   Basis. Contribution of Working Group 1 to the Fourth assesment report of
%   Intergovernmental Panel on Climate Change,}  S.~Solomon, D.~Qin, M.~Manning,
%   Z.~Chen, M.~Marquis, K.~B. Averyt, M.~Tignor, and H.~L. Miler, eds.
%   (Cambridge University Press, 2007).

% \end{thebibliography}

\end{document}